# A Database of Ultrastable MOFs Reassembled from Stable Fragments with Machine Learning Models

Aditya Nandy[1,2], Shuwen Yue[1], Changhwan Oh[1,3], Chenru Duan[1,2], Gianmarco Terrones[1],

Yongchul G. Chung[1,4], and Heather J. Kulik[1,2]*

[1]*Department of Chemical Engineering, Massachusetts Institute of Technology, Cambridge, MA 02139, USA*

[2]*Department of Chemistry, Massachusetts Institute of Technology, Cambridge, MA 02139, USA*

[3]*Department of Materials Science and Engineering, Massachusetts Institute of Technology, Cambridge, MA 02139, USA*

[4]*School of Chemical Engineering, Pusan National University, Busan 46241, Korea (South)*

*Correspondence: hjkulik@mit.edu

SUMMARY: High-throughput screening of large hypothetical databases of metal–organic frameworks (MOFs) can uncover new materials, but their stability in real-world applications is often unknown. We leverage community knowledge and machine learning (ML) models to identify MOFs that are thermally stable and stable upon activation. We separate these MOFs into their building blocks and recombine them to make a new hypothetical MOF database of over 50,000 structures that samples orders of magnitude more connectivity nets and inorganic building blocks than prior databases. This database shows an order of magnitude enrichment of ultrastable MOF structures that are stable upon activation and more than one standard deviation more thermally stable than the average experimentally characterized MOF. For the nearly 10,000 ultrastable MOFs, we compute bulk elastic moduli to confirm these materials have good mechanical stability, and we report methane deliverable capacities. Our work identifies privileged metal nodes in ultrastable MOFs that optimize gas storage and mechanical stability simultaneously.

## 1. Introduction.

Metal–organic frameworks (MOFs) have been the target of recent materials discovery efforts[1-3] due to their promise as catalysts[4,5] and functional materials[6]. Reticular MOFs structures are constructed from inorganic secondary building units (SBUs) and organic linkers, making them highly tunable.[7] For example, one may exchange metal atoms while maintaining SBU connectivity to tune catalytic properties[8] or electronic structure.[9,10] One may alternatively change the SBU connectivity while holding metal identity fixed to change MOF properties.[11-13] By varying linker identity and holding the SBU fixed, we can tune the pore size to alter gas diffusion and adsorption properties.[14] One may also decorate linkers with functional groups to influence gas adsorption selectivity[15] and catalysis.[16] Although these design knobs enable MOFs to be useful for gas storage[17,18], separations[3,19], and catalysis[20-26] or give them with novel properties that impart unique conductivity[27-30] and capabilities for sensing[29,31-33], these materials face limitations stemming from poor stability.[34,35] Many MOFs that are crystalline after synthesis either collapse upon activation or lack the thermal stability needed for practical applications.[36-38]

*In silico* databases play a key role in screening to identify promising materials for gas adsorption and separations.[3,39,40] Although lead compounds can be found in these databases, the activation and thermal stability of the MOFs is generally not considered during database construction. Instead, after computational screening has occurred, expert chemists analyze the best MOFs to attempt experimental synthesis.[3,40] There have been some efforts to compute the likely stability of these MOFs from simulations.[41] However, these simulations are expensive, cannot be carried out exhaustively, and may not faithfully describe experimental outcomes. Because it is challenging to predict the impact of including a novel building block in a MOF, *in silico* databases have been constructed with limited sets of well-studied building blocks. This *ad hoc* selection of MOF building blocks and nets in hypothetical databases has led to a bias in hypothetical MOF diversity.[42] Despite their relevance in catalysis, many transition metals that appear in experimental MOFs are entirely missing from hypothetical databases.[42,43]

The use of data-driven machine learning (ML) models has enabled the prediction of MOF metal oxidation states[44], band gaps[45], and gas adsorption[42,46-48] properties from experimental and computational data. We recently obtained measures of experimental stability using natural language processing (NLP) tools to mine the extant experimental literature.[34,49] We trained ML models that predict the experimental thermal and activation stability of MOFs based on their connectivity[42] and pore geometry.[50,51] This choice of features also permits application of the ML models to assess properties of MOFs in hypothetical datasets.

In this work, we use these ML models to identify experimentally known MOFs from an experimentally characterized database (i.e., CoRE MOF)[52] that are ultrastable. We develop a new approach to separate these MOFs into their building blocks and construct an *in silico* database from fragments of stable MOFs. Our database contains one and two orders of magnitude more building blocks and nets, respectively, than are present in existing *in silico* databases[3,39,40], thus enriching the diversity of transition metals in MOF nodes. This procedure enriches the database with MOFs that are likely to be thermally stable and stable upon activation by one order of magnitude. Because our *in silico* database MOFs contains nets that have not yet been sampled and could influence MOF mechanical stability in unexpected ways, we confirm that thermally stable and activation stable MOFs remain mechanically stable.[53-56] We demonstrate an application of this database by simulating the methane deliverable capacity of the resulting MOFs and identifying design rules that make a MOF stable (i.e., with respect to activation, thermally, and mechanically) while being suitable for methane adsorption.

## 2. Results and Discussion.

## 2a. Fragmenting and Representing Reticular Materials

Inorganic secondary building units (SBUs) and organic linkers are arranged in a net with a specific building block stoichiometry and connectivity to produce a MOF (Figure 1). Redundant definitions of what comprises SBUs make *in silico* MOF construction challenging. In contrast to organic linkers, which are well-defined, definitions of inorganic SBUs frequently but do not always contain fragments of the linkers (Figure 1). This "double counting" of the same atoms in both MOF components gives rise to challenges in *in silico* MOF recombination because one set of atoms from the SBU or linker must be removed. For example, UiO-66 is a commonly studied MOF that comprises a terephthalic acid linker and a Zr metal-oxyhydroxy cluster SBU, where terminal carboxyl groups are counted as part of both the SBU and the terephthalic acid linker (Figure 1). Developing a recombination strategy that overcomes these limitations is necessary for computational design of other materials, such as covalent organic frameworks (COFs) as well.

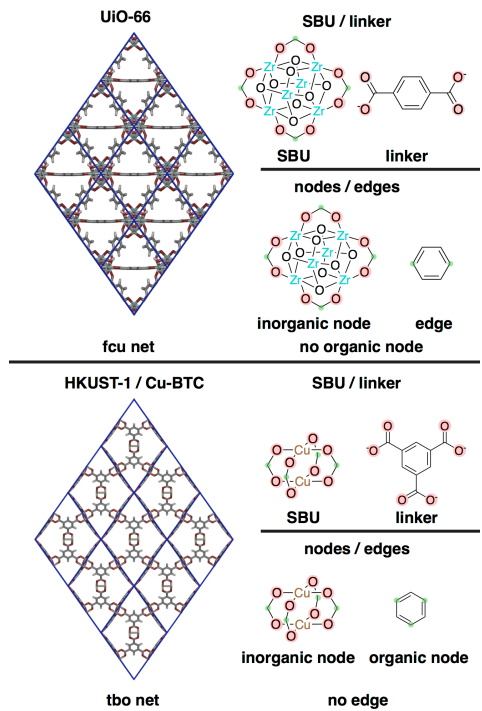

**Figure 1.** (top) The UiO-66 MOF, constructed from a Zr-oxyhydroxy cluster and terephthalic acid linkers in a *fcu* net. (bottom) The HKUST-1 MOF, also known as Cu-BTC, constructed of Cu paddlewheel SBUs and benzene tricarboxylic (BTC) acid linkers in a *tbo* net. Metal-coordinating oxygen atoms are highlighted with translucent red circles. Connection points for carbons are highlighted as green dots. Inset crystal structures are shown, with Cu in brown, Zr in blue, C in gray, O in red, and H in white. Unit cells are shown as blue boxes, with 3x3x3 super cells shown for both UiO-66 and HKUST-1.

A less ambiguous way to define a MOF structure from mutually exclusive building blocks (i.e., with no overlap in atoms between building blocks) is to consider the underlying connectivity of the MOF molecular graph. Here, we can think of the MOF as a set of nodes and edges pieced together in a graph. The node can be inorganic or organic, but the edge must be organic and have only two connection points that connect two nodes. From these mutually exclusive building blocks, we can reconstruct MOFs from fragments. Inorganic nodes are analogous to inorganic SBUs and are any building block containing a metal (Figure 1). Similarly, organic nodes are any organic components that have greater than two connection points. In contrast, organic edges are organic fragments that have two connection points. The combination of an inorganic node with an organic edge can give rise to many different linkers when the inorganic node has chemically distinct connection points (see Sec. 2b and Supporting Information Figure S1). In some cases, a MOF can lack an organic edge. The commonly studied HKUST-1 MOF has a benzene-tricarboxylic acid (BTC) linker that can be represented by a single inorganic node and a single organic node (Figure 1). Because there are no organic building blocks that form two connection points in this case, this MOF has no organic edge. By recombining inorganic nodes, organic nodes, and organic edges, we construct new MOFs from their reticular parts. The recombination strategy we take with nodes and edges is general for other materials, such as COFs, that have defined building blocks that are heuristically chosen.[57-59]

## 2b. Existing Hypothetical Databases Lack Ultrastable MOFs

Hypothetical MOF databases have typically been constructed with "tinker-toy" algorithms that recombine MOF nodes and edges to generate *in silico* databases for screening.[3,39,40] We assess three hypothetical databases that have previously been studied for gas adsorption and analyzed for their relative chemical diversity[42]: 1) hMOF[40], 2) BW-DB[3], and 3) ToBaCCo[39]. The databases are composed of tens-to-hundreds of thousands of structures with a limited number of inorganic nodes: hMOF contains 137,000 structures with five nodes, BW-DB contains 300,000 structures with eight nodes, and ToBaCCo contains 13,000 structures with 12 nodes. ML models trained on literature data[34,49] can be used to identify ultrastable hypothetical MOFs by using connectivity (i.e. revised autocorrelations, or RACs) and geometric features (e.g. pore size and surface area) as inputs.[34,49] We define ultrastability as thermal stability that is one standard deviation ($87°C$) more stable than the average CoRE MOF[49] (average decomposition temperature, $T_d = 359°C$) while simultaneously retaining activation stability (activation stability prediction $> 0.5$). If such hypothetical materials are ultrastable, then they would merit follow-up study by experiment or simulation, e.g., for applications in thermal catalysis where such stability is a necessity.

The hMOF database is composed of MOFs that are constructed from five inorganic nodes containing V, Cu, Zn, and Zr (Supporting Information Figure S2). As a consequence of the limited number of inorganic building blocks, this database contains six underlying net topologies (*pcu, sra, dia, tbo, nbo*, and *fcu*), where 90% of the structures have the *pcu* topology.[60] The hMOF set contains a large number (~137,000) of structures, and the MOFs in this database are less thermally stable on average compared to CoRE MOFs as predicted by our models (hMOF average $T_d$ = 337°C; CoRE MOF average[49] $T_d$: 359°C, Figure 2 and Supporting Information Figure S3). Although the majority (60%) of the MOFs are predicted to be stable upon activation, few demonstrate ultrastability (Figure 2 and Supporting Information Table S1). In total, less than 1% (i.e., 416) of the MOFs in the hMOF data set are ultrastable, and this number is further reduced when model uncertainty quantification[34,61] is employed to limit predictions to those that are strongly supported by the existing data (Supporting Information Figure S4).

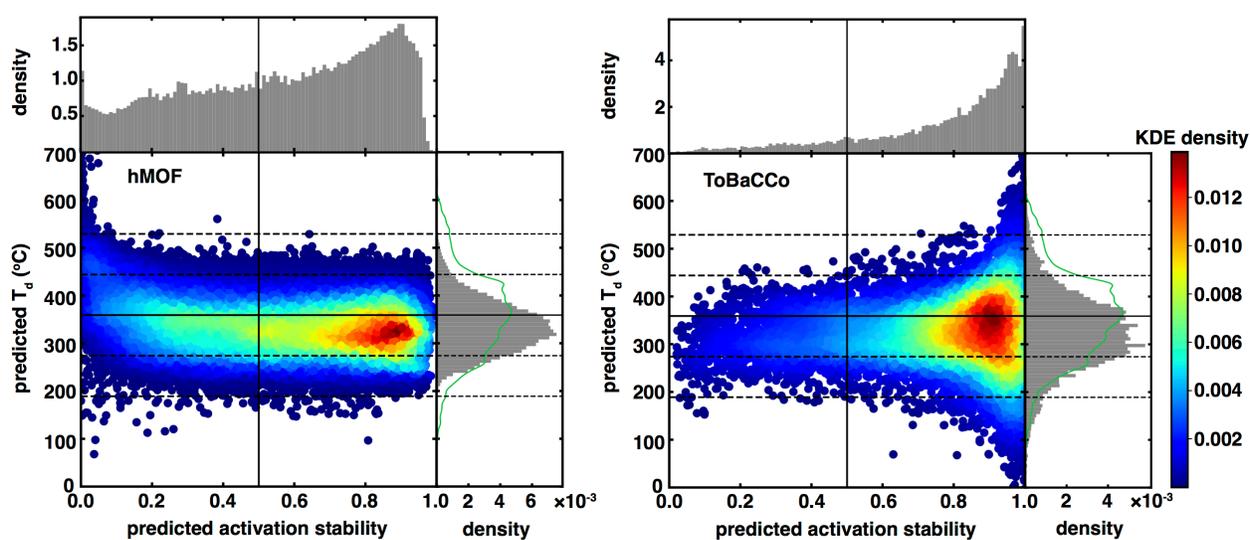

**Figure 2.** Predicted stability of members of the hMOF (left) and ToBaCCo (right) databases. ANN models from prior work are used to predict activation stability (unstable: 0 and stable: 1) and thermal decomposition temperature ($T_d$, °C). Data are colored by two-dimensional kernel density estimation (KDE) density values, as indicated by color bars. A solid black line on the x-axis indicates a value of 0.5 for activation stability. The solid black line on the y-axis indicates a $T_d$ of 359°C which is the average of the CoRE MOF training data. The dashed lines represent one (87°C) and two (174°C) standard deviations above and below the average of the CoRE MOF training data. Marginal histograms are shown in gray for both plots, with count densities on their corresponding axes. A 1D KDE plot of CoRE MOF $T_d$ values is shown as a green line for comparison to both thermal stability $T_d$ distributions.

In an effort to build a larger and more diverse hypothetical database for $CO_2$ separation materials, BW-DB was constructed from eight distinct inorganic nodes containing Al, V, Cr, Ni, Cu, and Zn (Supporting Information Figure S5). Together, with 12 distinct net topologies, this hypothetical database consists of more than 300,000 MOF structures (Supporting Information Table S2). Despite the increased number of structures and nodes sampled in this database, an exceedingly small (i.e., 767 or < 1%) number of the BW-DB MOFs are simultaneously stable upon activation and highly thermally stable (Supporting Information Figures S6 and S7). The average BW-DB MOF is less thermally stable than the average CoRE MOF (BW-DB average thermal decomposition temperature, $T_d$ = 335°C). As in our prior observations[34], we find that inorganic nodes influence thermal stability (i.e. average $T_d$: 296°C for a Cu paddlewheel and average $T_d$: 389°C for a V-oxo cluster). Nevertheless, thermal stability ranges (i.e., variation in $T_d$) for a single inorganic node can span over 300°C due to differences in organic edge or MOF net topology, indicating that judicious choice of organic edge should lead to the discovery of more stable MOFs (Supporting Information Figure S8).

In comparison to hMOF and BW-DB databases, ToBaCCo[39] samples a greater number of underlying net topologies (41 in total) to enrich MOF net diversity. ToBaCCo accomplishes this diversity while only sampling a subset of metals in the SBUs, namely Mn, Co, Cu, Zn, and Zr (Supporting Information Figure S9). Nevertheless, even though ToBaCCo is an order of magnitude smaller in size than hMOF or BW-DB (~13,000 structures), it contains both a larger number (i.e.,

750) and percentage (i.e., 6.5%) of MOFs that are predicted to be simultaneously stable upon activation and highly thermally stable (Figure 2). Despite larger tails for the thermal stability distribution, the average MOF in ToBaCCo is also less thermally stable than CoRE MOFs (ToBaCCo average thermal decomposition temperature, $T_d = 324°C$). Indeed, we find that certain MOF topologies absent from the other two hypothetical MOF sets are enriched among the ultrastable ToBaCCo MOF subset. As an example, the *csq* topology appears in 603 out of 11,754 (5%) MOFs over the full space but is enriched in the ultrastable subset, with 226 out of 750 (30%) MOFs exhibiting this topology. Therefore, the incorporation of novel and less-explored MOF topologies may enable the design of stable MOFs, although it should be noted that the diversity of the ToBaCCo MOFs relative to the CoRE MOF training data results in many fewer of the predictions being strongly supported according to our uncertainty quantification criteria (Supporting Information Figure S10).

When we use our criterion for identifying MOFs that are stable upon activation and highly thermally stable, many of the inorganic nodes that are associated with high stability in CoRE MOF are absent from the current hypothetical databases. The lack of ultrastable MOFs, as especially evident in hMOF and BW-DB, along with their limited metal diversity motivates the construction of new databases that are enriched with stable MOFs. Given the higher diversity of inorganic nodes along with the evidence of their synthetic feasibility and stability, CoRE MOF is an ideal source of building blocks. The improved stability of ToBaCCo MOFs due to increased net diversity demonstrates the need to study and sample a wider range of nets. We thus devise a strategy to recombine building blocks mined from the CoRE MOF database with increased net diversity to access areas of improved MOF stability.

**2c. Constructing New MOFs from Building Blocks Mined from Stable MOFs**

To use CoRE MOF as a source of inorganic nodes to generate a new dataset, we adapted our previous strategy of mining SBUs[42,49] to ensure that all inorganic nodes can be recombined to construct new MOFs (see Experimental Methods). With the CoRE MOF dataset, we identify the MOFs that have either 1) been identified as highly thermally stable[34,49] (>1 standard deviation above the mean $T_d$) and stable upon activation according to the published article tied to the MOF or 2) predicted by our models with high confidence[34,49,61] to be highly thermally stable and stable upon activation. Then, we mine and recombine their building blocks to construct novel but likely synthesizable MOFs that we re-evaluate with our ML models.

Using our criteria for ground truth or model-predicted stability, we identify a stable subset of 384 CoRE MOFs that we mine for building blocks (Supporting Information Table S3). Analysis of this subset of MOFs demonstrates that metals that are associated with stability (e.g., Zr) are missing (Figure 3 and Supporting Information Figure S11). This omission could be due to limited literature reports or due to their absence in the original CoRE MOF subset we used (i.e., CoRE MOF 2019 all solvent removed or ASR). To address this limitation, we curated a set of MOFs that was not deposited with the original CoRE 2019 data set, which we call the "extended CoRE 2019 data set" (Supporting Information). This data set contains structures that were not sanitized (i.e., with solvent molecules removed) and thus not deposited with the other CoRE MOFs. By repeating our text mining analysis and applying our models with uncertainty control, we analyzed 2,123 candidate MOFs and identified 90 additional stable MOFs (Supporting Information Table S4 and Figure S12). This additional set increases the diversity of stable building blocks even further by adding 11 new elements (including Zr) that were not present in the original set (Figure 3 and Supporting Information Figures S11 and S12). The extended CoRE MOF 2019 data set is also a natural test set with unseen MOFs for our ML models. We confirm on this new set aside test set that we achieve ML model prediction errors comparable to that of the original test set (i.e., $T_d$ MAE of 44 °C) by making predictions on points with low uncertainty (Supporting Information Figure S13). We combine these two sets together to obtain 474 stable MOFs from which to mine building blocks.

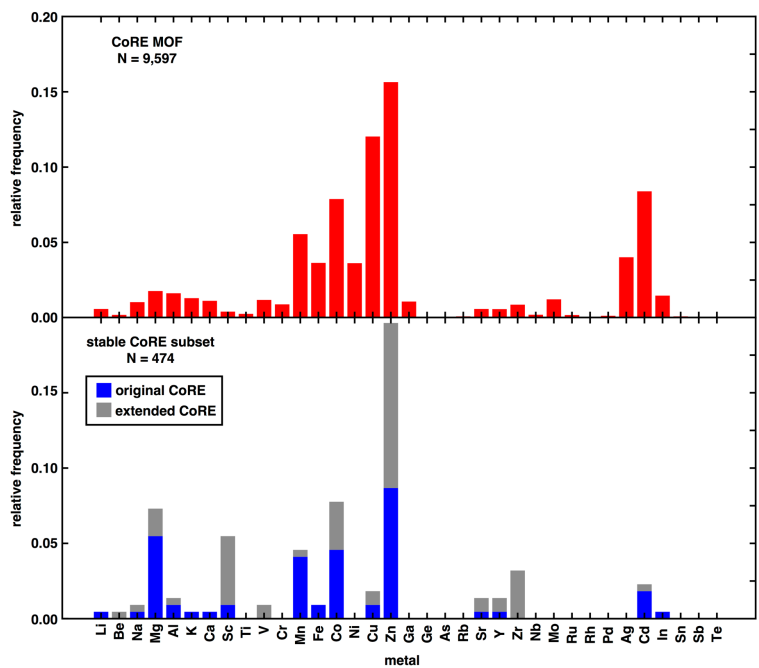

**Figure 3.** (top) Bar plots of the 3d, 4d, and 5d metals present in the full CoRE MOF dataset. (bottom) Stacked bar plots of the 3d, 4d, and 5d metals present within the stable subset of the original and extended CoRE MOF subsets. All y-axes are shown as relative frequencies, and metals labeled by their two-letter abbreviations.

We also identify organic nodes and edges from the stable MOFs to recombine with the inorganic nodes (Figure 4 and Supporting Information Figure S14). We developed an automated strategy to identify and avoid 1D rod inorganic nodes that cannot easily be truncated and used for MOF construction (Supporting Information Figure S15). A significant fraction (37%) of the mined inorganic nodes from stable MOFs are 1D rods, with the remaining nodes being finite and suitable SBUs for reconstruction (Figure 4). Analysis of the mined organic building blocks from stable MOFs indicates that edges with non-carbon or hydrogen heteroatoms are uncommon (3 out of 16, 19%) among the stable subset (Supporting Information Figure S14). While identifying organic edges, we avoid cases that would introduce a geometric mismatch[62] when recombining nodes and edges (see Experimental Methods). For each inorganic node identified to be unique by connectivity[63], we measured the compatibility of the node with each of the possible nets and selected the structure of the building block that was compatible with the most nets (Supporting Information Figure S16). The final set of building blocks corresponds to 88 inorganic nodes, 32 organic nodes, and 16 organic edges to reconstruct new MOFs (Supporting Information).

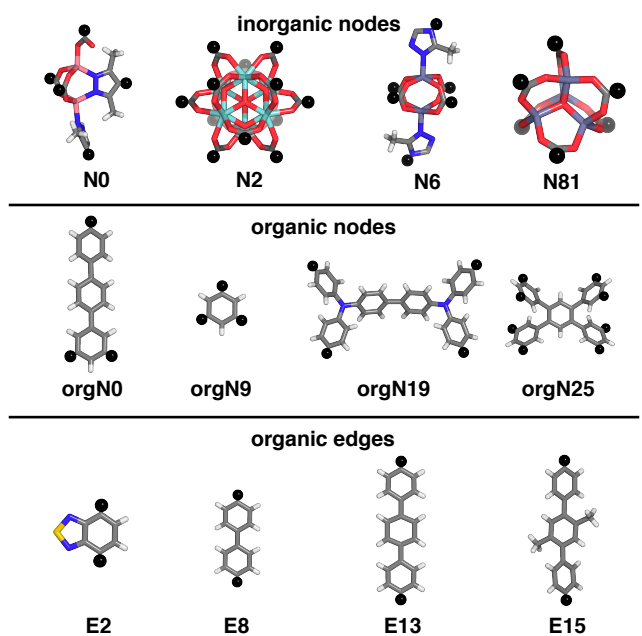

**Figure 4.** (top) Representative inorganic nodes (top), organic nodes (middle), and organic edges (bottom) extracted from stable MOFs. Atoms are colored as follows: white for H, gray for C, blue for N, red for O, yellow for S, pink for Co, purple for Zn, teal for Zr, and black for X, which represents the connection points of the node or edge respectively.

With our *de novo* strategy, metal diversity, measured by the unique metals used in inorganic nodes, increases 5-fold for our building blocks relative to the building blocks in most diverse hypothetical database considered (e.g., BW-DB). Comparison of unique inorganic node connectivities (i.e., differentiating between a Zn paddlewheel and Zn trimer) reveals a 10-fold increase in inorganic node diversity relative to the building blocks present in existing databases.

Moreover, after evaluating compatibility between building blocks and connectivity nets, we use 755 distinct connectivity nets for MOF construction, which is a 20-fold increase from the number of nets (41) present in the topologically diverse ToBaCCo MOF set. In addition, 88% (28 out of 32) of the organic nodes and 50% (8 out of 16) of the organic edges are do not appear in the large hypothetical databases.

We first consider inorganic node–organic node–organic edge configurations of stable MOFs in the CoRE MOF set to inspire the construction of our hypothetical database. For the CoRE MOFs that are stable upon activation and have above average thermal stability, 88% of MOFs consist of one of the following: 1) one inorganic node and one edge (63%), 2) one inorganic node, one organic node, and one edge (16%), or 3) two inorganic nodes and one edge (9%). Because these three configurations capture most MOFs, we thus construct our new set of MOFs based on these three building block configurations. We use the tinker-toy algorithm in PORMAKE[48] to construct 109,995 total MOFs. After optimizing these MOFs with a 20-minute wall time constraint using our structural optimization algorithm (see Experimental Methods), we have a final data set of 54,139 MOFs that we can evaluate using our ML models (Figure 5 and Supporting Information Table S5).

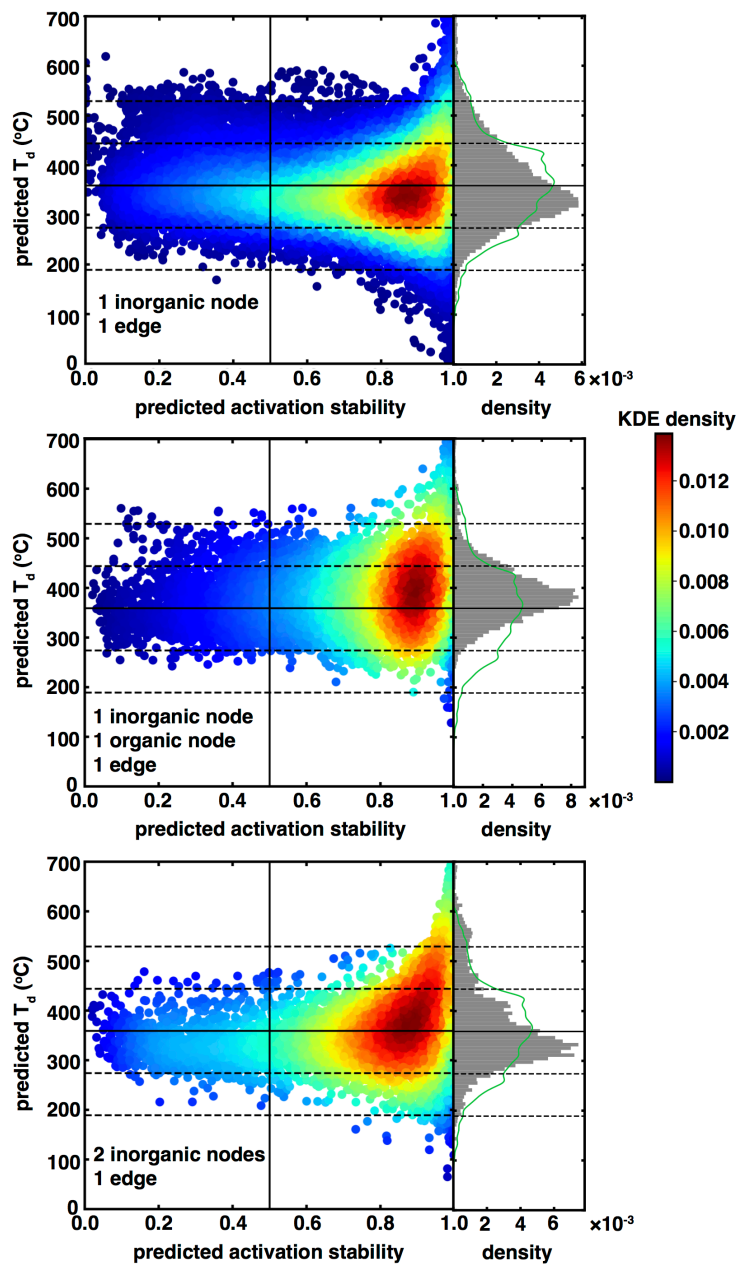

**Figure 5.** Predicted stability of the ultrastable MOF database, grouped by connectivity type. ANN models from prior work are used to predict activation stability (unstable: 0 and stable: 1) and thermal decomposition temperature ($T_d$, °C). Data are colored by two-dimensional kernel density estimation (KDE) density values, as indicated by color bars. A solid black line on the x-axis indicates a value of 0.5 for activation stability. The solid black line on the y-axis indicates a $T_d$ of 359°C which is the average of the CoRE MOF training data. The dashed lines represent one (87°C) and two (174°C) standard deviations above and below the average for the CoRE MOF training data. Marginal histograms are shown in gray for both plots, with count densities on their corresponding axes. A 1D KDE plot of CoRE MOF $T_d$ values is shown as a green line for comparison to both thermal stability $T_d$ distributions.

When we use the ML models to evaluate the stability of our newly constructed MOF database, we find that nearly half (47%, 25,336 out of 54,139) of the MOFs are predicted to be thermally stable and stable upon activation. We find that 36% of the MOFs (19,342 out of 54,139) are predicted to be stable upon activation but have below-average thermal stability. A significantly smaller percentage (8%, 4,285 out of 54,139) are predicted to be thermally stable but not stable upon activation (Supporting Information Table S6). When considering the ultrastable stability criteria (i.e., activation stability > 0.5 and > 1 standard deviation above the mean for thermal stability), we observe an order of magnitude enrichment (18%, 9,524 out of 54,139) in ultrastable MOFs relative to other hypothetical databases. Notably, this ultrastable subset contains at least one example of 517 (68% of 755) of the nets used for initial construction while only 10% (77 out of 755) of the nets only appear in both thermally and activation unstable MOFs. We find that the organic edges E8 and E13 (i.e., the edges used in UiO-67 and UiO-68 respectively) and the phenylated and methylated derivatives of E13 (i.e., E1 and E15) are enriched in stable MOFs (Supporting Information Figure S14 and Table S7). Even these edges, however, can lead to unstable MOFs when combined with specific inorganic nodes. Although the organic edge E13 is enriched in stable MOFs overall, when combined with inorganic node N0, it tends to make more unstable MOFs, depending on the net (unstable MOFs with N0 and E13: 80, ultrastable MOFs with N0 and E13: 49). The enriched stability by MOF building block combination type, however, suggests that MOFs with two inorganic nodes and one edge are enriched in stable MOFs (Figure 5). Therefore, MOFs with more than a single node should be of increased focus to enrich the design of ultrastable MOFs. This finding contrasts with the focus of existing MOF databases on MOF topologies comprising one inorganic node and one organic edge.

**2d. Screening Thermally and Activation Stable Hypothetical MOFs for Mechanical Stability and Methane Deliverable Capacity**

Because we are constructing an *in silico* database of MOFs with nets that have not been explored extensively, we sought to confirm that our MOFs are mechanically stable. This evaluation is important because MOFs that are useful for practical applications must not only be thermally stable and amenable to activation but must also retain structural integrity under mechanical force. During gas adsorption, MOFs must be able to avoid structural distortion arising from hydrostatic compression[64] that can lead to pressure-induced amorphization.[65] Similarly, during catalysis, MOFs must be able to withstand large thermal gradients and fluxes of reactants and products, requiring MOFs to be both thermally and mechanically stable.[64] Previous studies that have focused on MOF mechanical stability have neglected thermal stability and activation stability due to an inability to predict these properties.[54,55] We use our ML models and hypothetical dataset construction approach to search for MOFs that simultaneously satisfy our ultrastability criteria (i.e., highly thermally stable and stable with respect to activation[34]) along with mechanical stability. For these MOFs, we study methane adsorption as a value-added property to analyze tradeoffs between different facets of stability and gas adsorption.

We compute the mechanical properties (i.e., bulk elastic and shear moduli) for the set of MOFs that models confidently predict to be highly stable with respect to the other two properties (see Sec. 2b and Supporting Information Table S8 and Text S1). While the intuitive dependence on pore size (i.e., longer edges leading to weaker materials) of mechanical properties has been previously observed[54], we now find dependence of the elastic and shear moduli on the identity of the inorganic node (Figure 6 and Supporting Information Figures S17 and S18). In accordance with prior conclusions on edge length, we find that increasing edge length while holding the inorganic node fixed reduces the elastic modulus (shorter E8: 20.2 GPa vs longer E1: 8.4 GPa). Altering the identity of the node from Co-based N39 to Gd-based N43 increases the elastic modulus 10-fold (N39: 2.6 GPa, N43: 20.2 GPa) even while holding the *wir* net fixed (Figure 6). Despite the overall reduction in mechanical stability with a longer edge, the MOF with the Gd-based N43 node is still four times more mechanically stable than the same MOF with the Co-based N39 node (Figure 6). Together, these analyses highlight the simultaneous importance of the inorganic node and organic edge in governing mechanical stability.

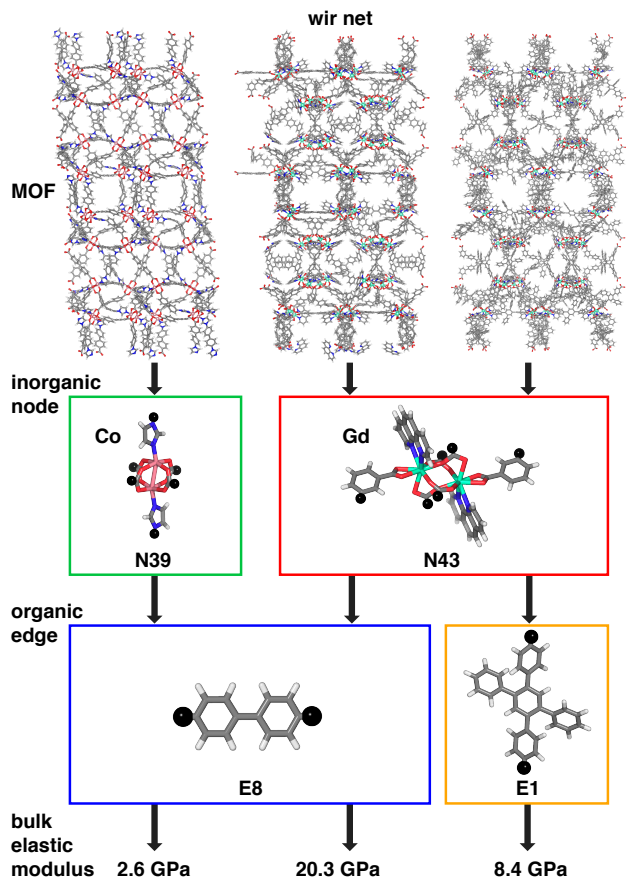

**Figure 6.** Three representative MOFs in the *wir* connectivity net showing variation from the Co-based inorganic node N39 to the Gd-based inorganic node N43, and edge E8 to edge E1. The resulting bulk elastic modulus of each combination of inorganic node and organic edge in a *wir* net is shown. Atoms are colored as follows: white for H, gray for C, blue for N, red for O, pink for Co, blue-green for Gd, and black for X, which represents the connection points of the node or edge.

Beyond the roles of individual nodes and edges, we find that MOFs containing an organic node are about 50% less mechanically stable on average relative to MOFs that do not contain one (Supporting Information Figure S19). Similarly, MOFs constructed from two inorganic nodes are not more mechanically stable relative to MOFs constructed from a single inorganic node, suggesting that their promise as ultrastable MOFs would have to be weighed against both lower mechanical stability and the more complex synthesis routes that would be required to achieve these MOF topologies (Supporting Information Figure S19). Conclusions on elastic modulus generally extend to shear modulus because the two properties are correlated (Pearson's $r = 0.88$, Supporting Information Figure S20). Analysis of a diverse set of MOFs that our models predicted to be thermally unstable and unstable with respect to activation to confirm that our mechanical stability observations were not biased by a focus on high thermal/activation stability MOFs (Supporting Information Figures S21 and S22). Considering only the ultrastable portion of the data set (9,524 out of 54,139 MOFs, or 18%) has minimal influence on the ranges of elastic moduli (Supporting Information Figure S22).

Due to the expected competing influence of linker length (i.e., from the substituent edge and organic nodes) on methane uptake and mechanical stability, we analyze the tradeoff for designing a mechanically stable MOF with reasonable methane deliverable capacities for MOFs that are thermally stable and stable with respect to activation. Among these compounds, MOFs with longer edges have higher methane deliverable capacities, likely due to increased pore sizes (Supporting Information Figure S23). Although many MOFs with two inorganic nodes are simultaneously thermally stable and stable upon activation, they are rarely mechanically stable (i.e., maximum bulk elastic modulus of 27 GPa), limiting motivation for further consideration of this type of MOF (Figure 7). The best tradeoffs occur for MOFs with one inorganic node and one edge, with 18 out of 25 Pareto-optimal MOFs coming from this class (Figure 7 and Supporting Information Table S9). For the remaining seven MOFs on the Pareto front that also contain an organic node, all have reduced elastic moduli and are at one extreme of the Pareto front (Figure 7 and Supporting Information Table S9). The most mechanically stable MOFs in this work have bulk elastic moduli that are comparable to mechanically stable MOFs from prior studies.[54] Finding MOFs with bulk elastic moduli of >50 GPa, however, remains a challenge when simultaneously considering methane deliverable capacity, because MOFs with smaller pores tend to be more mechanically stable[54] (Figure 7).

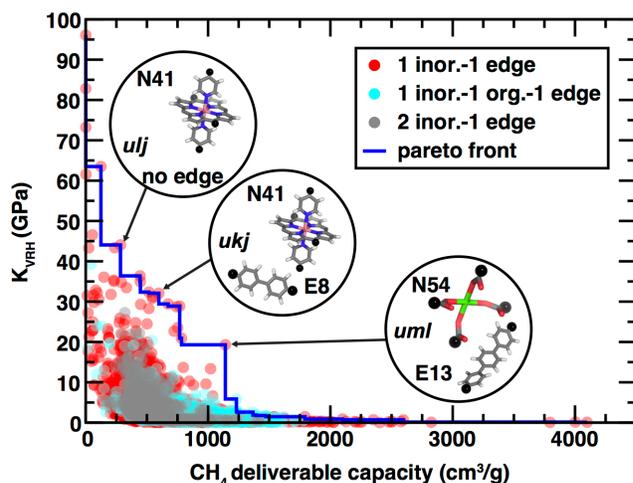

**Figure 7.** The relationship between gravimetric methane deliverable capacity (in cm$^3$/g) and elastic modulus (K$_{VRH}$, in GPa) for all thermally and activation stable MOFs. Each connectivity type is shown in a different color: MOFs composed of one inorganic node and one edge in red, one inorganic node, one organic node, and one edge in blue, and two inorganic nodes and one edge in gray. Data are shown as translucent circles to depict overlap. A blue line indicates the Pareto front for these two properties, with three representative examples showing MOFs on the Pareto front and their constituent nets, inorganic nodes, and organic edges. Atoms are colored as follows: white for H, gray for C, blue for N, red for O, green for Mg, pink for Co, and black for X, which represents the connection points of the node or edge.

We find that E13, [1,1′:4′,1-terphenyl]-4,4″-dicarboxylic acid or TPDC, is frequently sampled (in 10 out of 25 MOFs) on the Pareto front in combination with nets that are absent from prior hypothetical databases, despite being a longer linker that leads to larger-pore MOFs (Supporting Information Table S9). The well-studied UiO-68 MOF contains TPDC in an *fcu* net and is typically unstable upon activation.[66] Yet our findings show that the edge that comprises TPDC should lead to a MOF that is simultaneously ultrastable (i.e., highly thermally stable and stable with respect to activation) and mechanically stable when present in more highly connected nets. Similarly, we find that the Co-based porphyrinic N41 inorganic node is common (i.e., 11 out of 25 examples) among the materials on the Pareto front and demonstrates an optimal tradeoff between mechanical stability and methane deliverable capacity. This is likely due to the rigid nature of the porphyrinic node. For the 14 remaining cases, the majority (i.e., 8 out of 14) use lanthanide or actinide nodes. The dominance of the Co-based N41 node and lanthanide- or actinide-containing nodes suggests they are useful candidate building blocks for MOFs with optimal tradeoffs between mechanical stability and methane deliverable capacity while retaining other measures (i.e., thermal and activation) of stability (Supporting Information Table S9). Longer aromatic edges such as E8 and E13 are enriched in MOFs that are stable upon activation, thermally stable, and mechanically stable (Figure 7 and Supporting Information). This suggests that focusing on these longer aromatic edges during MOF construction in higher connectivity nets can likely produce MOFs that are ultrastable, in contrast to the focus of experimental efforts on smaller linkers such as benzene-dicarboxylic acid. While these identified MOFs represent leads themselves, these combinations of nodes and edges could also provide starting points for subsequent property optimization, e.g., when paired with linker functionalization strategies that could further increase the number of candidate materials with optimal properties with even modest trade-offs in stability expected to still yield stable materials. Beyond MOFs, combinations of edges and organic nodes can also lead to the discovery of other reticular materials, such as zeolites or COFs, with a set of building blocks that rely less on heuristics and thus have enhanced diversity.

Indeed, previous work on COFs has demonstrated that the consideration of increased net diversity leads to novel materials with promising properties, only two of which corresponded to previously synthesized materials.[57] The stability of the structures, however, remained in question, since there were no models to evaluate which compounds are most likely to be stable in that case.

## 3. Conclusions

The tailored active sites and pores in MOFs make them promising candidate materials for an array of applications, motivating the *in silico* screening of databases of combinatorially assembled MOF structures. A lack of consideration of stability, however, limits the experimental realization of computationally-predicted MOFs. This same lack of consideration of stability also limits the realization of other predicted materials, such as COFs. Building blocks used to construct databases for screening are typically selected by hand with domain knowledge, leading to bias in the metals and connectivity nets present and a lack of diversity in comparison to experimentally synthesized MOFs. Applying our machine learning models, trained to predict experimental data on thermal and activation stability on such databases, confirmed that most hypothetically-constructed MOFs are unlikely to have the same stability achieved by experimentally synthesized MOFs.

We developed an approach to fragment MOFs into their building blocks and recombine fragments of stable MOFs to construct a diverse database. With our *de novo* reconstruction strategy, our database samples two orders of magnitude more connectivity nets than existing databases and an order of magnitude more metals than existing hypothetical databases, thus better reflecting the diversity present in experimental databases. These novel combinations retained the good characteristics of the parent MOFs: we showed an order of magnitude enrichment in ultrastable (i.e., highly thermally and activation stable) MOFs in this database in comparison to prior databases. We observed a strong dependence on linker chemistry for the MOFs that are the most stable, finding that aromatic hydrocarbon edges are enriched in stable MOFs. An analogous approach is applicable to other reticular materials, such as COFs, that are also typically constructed from heuristically chosen building blocks.

From our database, we screened all thermally and activation stable materials for their mechanical stability and found that our MOFs generally retained good mechanical stability. In real-world applications it is common to screen for a target property that is expected to have inherent tradeoffs against common measures of stability. As a demonstration of the use of our database we considered the relationship between methane deliverable capacities and mechanical stabilities, two properties expected to have opposing relationships with porosity, from the set of thermally stable MOFs that are also stable upon activation. MOFs containing Co-porphyrin, lanthanide, or actinide inorganic nodes in combination with more highly connected nets and longer linkers have the best tradeoff for remaining thermally stable, stable upon activation, and mechanically stable while adsorbing methane, counter to expectations of longer linkers universally weakening mechanical stability. By considering experimental stability during *in silico* database construction, our database of ultrastable MOFs will accelerate the discovery of new, practically usable MOFs that can be used for high temperature applications. This strategy is expected to be beneficial for other porous materials (e.g., COFs) that can be designed from the ground up from their building blocks.

**Experimental Procedures**

**4a. MOF Building Blocks and Construction.**

We study periodic experimental MOF structures that correspond to previously studied[34,49] MOFs obtained from a subset of the all solvent removed (ASR) portion of the CoRE MOF 2019 dataset.[52] Using this set of MOFs, we break down their structures into constituent inorganic nodes, organic nodes, and organic edges for reconstruction of new MOFs (Supporting Information Figure S15). We define any metal-containing building block as an inorganic node, any organic building block that forms two connections with other units as an edge, and any organic building block with more than two connection points as an organic node (Supporting Information Figures S15 and S24). We use molSimplify[67] to separate MOFs into their constituent building blocks, assign building block connectivity within an .XYZ file, and identify connection points between building blocks (Supporting Information Figure S24). In order to place compatible building blocks for each net, we used predefined routines in PORMAKE[48] to measure the root-mean-square deviations (RMSDs) of inorganic and organic node building blocks to MOF net templates. We use a suggested 0.3 Å threshold[48] to determine if a node is compatible with a net. Some organic edges have geometries that make them incompatible with most nets[62] because the vectors of the carbon atom to connection point bonds are not collinear. We selected heuristically a cutoff of 150° for this transverse angle to choose edges compatible with MOF reconstruction (Supporting Information Figure S25).

We append the .XYZ file format with extra records containing the non-periodic building block connectivity to construct and store the periodic graph for the MOF. We also append the periodic graph connectivity in the .CIF file. We use this graph in the LAMMPS interface[68] to prepare all input files for energetic minimization using the UFF4MOF force field.[69,70] We optimize all MOF structures with this force field using LAMMPS for up to 20 minutes.[71] For this step, we modify a previously reported procedure for structural optimization[57] but allow unit cell angles and lengths to change with all other defaults applied (Supporting Information Figure S26).

## 4b. Machine Learning Features.

We compute revised autocorrelations[42,72] (RACs) on periodic molecular graphs for all MOFs using molSimplify.[67] RACs are products and differences of five atom-wise properties: nuclear charge ($Z$), identity ($I$), topology ($T$), electronegativity ($\chi$), and covalent radius ($S$) evaluated over a bond difference, $d$, with a maximum depth of 3. RACs are centered on the metal (metal-centered RACs), centered on linker coordinating atoms (linker-centered RACs), evaluated over the full linker (full-scope linker RACs), or evaluated over the full unit cell (full-scope RACs). If a MOF has multiple metal atoms, metal-centered RACs are evaluated over each metal and averaged by the number of sites. Any non-carbon or non-hydrogen heteroatom that is part of the linker but is not metal-coordinating is defined as a functional group. There are 134 RAC features after eliminating constant features (Supporting Information Table S10). For the MOFs constructed using PORMAKE[48], we use the graph connectivity stored in the .CIF file. For the other hypothetical MOF databases (hMOF[40], BW-DB[3], and ToBACCo[39]), we interpret the graph from the atomic coordinates in the .CIF file using molSimplify.[67] We computed geometric properties (e.g., pore volume) with Zeo++, using a nitrogen probe molecule[50,51] with a 1.86 Å radius, resulting in 14 geometric features (Supporting Information Table S11). All models trained in prior work[34,49] used these RACs and geometric features as inputs. Prior to model predictions, all features were scaled based on the model training data.

## 4c. Molecular Simulation for Mechanical Properties.

All mechanical properties were calculated using the LAMMPS molecular simulation package.[71] We extracted the moduli of elasticity from the 6x6 stiffness matrix.[73] This tensor contains all information about the mechanical properties of a material in the elastic regime of the stress-strain curve. We calculate the stiffness matrix by applying a maximum strain of 1% and evaluating the relative energy difference between the deformed structure and the initial structure. The initial structures used for mechanical property calculations had optimized and relaxed cells as described in Sec. 2a. All mechanical property calculations used the UFF4MOF[69,70] force field with conjugate gradient minimization for geometry optimization.

**4d. Molecular Simulation for Methane Deliverable Capacity.**

We performed grand canonical Monte Carlo (GCMC) simulations using the RASPA simulation package[74] to obtain methane uptake at 298 K and two pressures, 65 bar and 5.8 bar, corresponding the ARPA-E target methane deliverable capacity standard.[17] We assumed rigid frameworks from the optimized hypothetical MOF structures and modeled guest–guest and host–guest interactions using a Lennard-Jones potential truncated and shifted at 12.8 Å. The force-field parameters for all framework atoms were taken from the UFF[69] force field. Methane molecules were represented by the united-atom TraPPE force field.[75] All cross interactions were calculated using Lorentz–Berthelot mixing rules.[76,77] Each unit cell was replicated in all dimensions to achieve a minimum cell length which exceeds twice the Lennard-Jones cutoff in order to avoid periodic effects. GCMC simulations were performed for 5000 initialization cycles and 5000 production cycles.

**Resource availability**

*Lead contact*

Further information and requests for resources and materials should be directed to and will be fulfilled by the lead contact, Professor Heather J. Kulik (hjkulik@mit.edu)

*Materials availability*

This study did not generate new, unique reagents.

*Data and code availability*

The scripts used for MOF construction can be found in the molSimplify GitHub repository: https://github.com/hjkgrp/molSimplify. Data associated with the ultrastable MOF database is hosted via Zenodo and has the following permanent DOI: 10.5281/zenodo.7091192. All data associated with this work is made publicly available including stability predictions and uncertainty quantification (UQ) values for hMOF, BW-DB, ToBaCCo, and the Ultrastable MOF database; all building blocks including inorganic nodes, organic nodes, and organic edges; all structures for the Ultrastable MOF dataset; all structures for the extended CoRE MOF dataset; computed mechanical properties and methane adsorption properties.

**Supporting Information**.

Example of nodes and edges that lead to multiple linkers; algorithm for fragmenting MOF into building blocks; representative building blocks for MOF reconstructions; distribution of connecting atom angles for edges used for MOF reconstructions; methodology for optimizing MOFs in the database; details and accounting for RACs and geometric features; inorganic nodes used in the hMOF database and corresponding MOF counts; distributions of thermal stability for hMOF database and the influence of uncertainty quantification; inorganic nodes used in the BW-DB database and corresponding MOF counts; distributions of thermal stability for BW-DB database and the influence of uncertainty quantification; inorganic nodes in the ToBaCCo MOF database; counts for stable MOFs within the CoRE MOF database; bar plot of metals present in the CoRE MOF database; details of extended CoRE MOF data set; edges used for MOF reconstructions; details of RMSD measurements and net compatibilities for building blocks; relative frequencies of linkers that appear in the ultrastable set; convergence statistics and details for mechanical property calculations; box and whisker plots of inorganic node and organic edge dependence for mechanical stability; box and whisker plots for mechanical stability of MOFs that contain different connectivities; comparison of bulk elastic and shear moduli; principal component analysis for thermally and solvent removal unstable MOFs; box and whisker plots of edge dependence of methane deliverable capacity; distributions of deliverable capacity values for stable and unstable MOFs; Pareto optimal MOFs for deliverable capacity and mechanical stability (PDF)

ACKNOWLEDGMENT

The authors acknowledge primary support for MOF construction by DARPA under grant numbers D18AP00039 (A.N. and C. D.). A.N. was partially supported by a National Science Foundation Graduate Research Fellowship under Grant #1122374. Algorithm and website development was supported by the Office of Naval Research under grant number N00014-20-1-2150 (C.D., G.T., and H.J.K.). This work is also partially supported as part of the Inorganometallic Catalysis Design Center, an Energy Frontier Research Center funded by the U.S. Department of Energy, Office of Science, Basic Energy Sciences under Award DE-SC0012702 (A.N.). This work is also partially supported as part of the Center for Enhanced Nanofluidic Transport, an Energy Frontier Research Center funded by the U.S. Department of Energy, Office of Science, Basic Energy Sciences under Award DE-SC0019112 (S.Y.). This work is also partially supported by an MIT Portugal Seed Fund (C.O.). This work was carried out in part using computational resources from the Extreme Science and Engineering Discovery Environment (XSEDE), which is supported by National Science Foundation grant number ACI-1548562. This work also made use of Department of Defense HPCMP computing resources. H.J.K. holds a Career Award at the Scientific Interface from the Burroughs Wellcome Fund, an AAAS Marion Milligan Mason Award, and an Alfred P. Sloan fellowship, which supported this work. The authors thank Adam H. Steeves for providing a critical reading of the manuscript.

**Author Contributions:**

Conceptualization, A.N. and H.J.K.; Methodology, A.N., S.Y., C.D., and H.J.K.; Investigation, A.N., S.Y., C.O.; Writing – Original Draft, A.N. and H.J.K.; Writing – Review & Editing, A.N., S.Y., C.O., C.D., and H.J.K.; Funding Acquisition, H.J.K.; Resources, A.N., G.T., Y.G.C., and H.J.K.; Supervision, H.J.K.

**Declaration of Interests**

The authors declare no competing financial interests.

**Supporting Information for**

***A Database of Ultrastable MOFs Reassembled from Stable Fragments with Machine Learning Models***

Aditya Nandy[1,2], Shuwen Yue[1], Changhwan Oh[1,3], Chenru Duan[1,2], Gianmarco Terrones[1], Yongchul Chung[1,4], and Heather J. Kulik[1,2]*

[1]*Department of Chemical Engineering, Massachusetts Institute of Technology, Cambridge, MA 02139, USA*

[2]*Department of Chemistry, Massachusetts Institute of Technology, Cambridge, MA 02139, USA*

[3]*Department of Materials Science and Engineering, Massachusetts Institute of Technology, Cambridge, MA 02139, USA*

[4]*School of Chemical Engineering, Pusan National University, Busan 46241, Korea (South)*

*email: hjkulik@mit.edu

**Contents**

**node:** N0, **edge:** E1

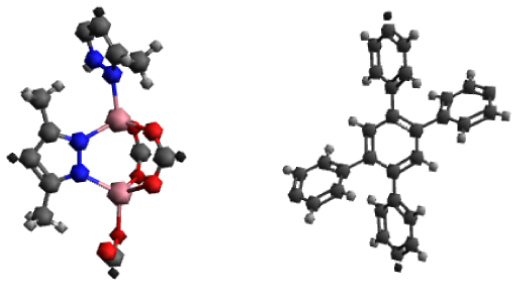

**linkers made in fee net**

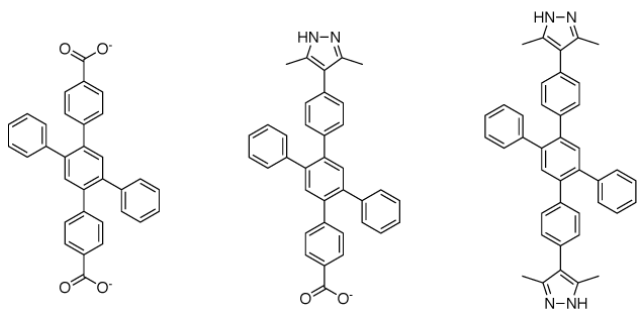

**Figure S1.** A representative example of an inorganic node and an organic edge that can lead to multiple different linkers when combined. In this example, we have a node (N0) combined with an edge (E1). Because the inorganic node has both an O- and N-coordinating moiety, it forms linkers with both of these metal-coordinating motifs.

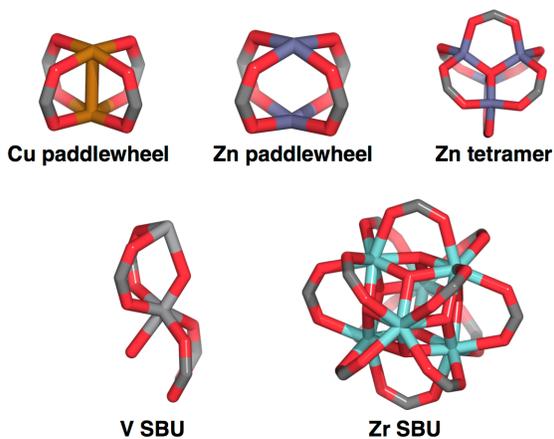

**Figure S2.** The five inorganic nodes used for all MOFs in the hMOF database. The atoms are colored as follows: red for O, gray for C, brown for Cu, purple for Zn, light blue for Zr, and light gray for V.

**Table S1**. Counts for the structures in the hMOF database that can be featurized and are predicted to be in the ultrastable set. We define the ultrastable set as MOFs that have a predicted activation stability value >0.5 and a predicted thermal stability that is one standard deviation (87°C) above average (359°C). Stability predictions are made on the featurizable subset of MOFs.

| step | MOFs remaining |
|---|---|
| start | 137,953 |
| featurized with RACs and geometric features | 114,646 |
| activation stability > 0.5 | 69,256 |
| thermal stability > (359+87°C) | 4,089 |
| (activation stability > 0.5) & (thermal stability > (359+87°C)) | 416 |

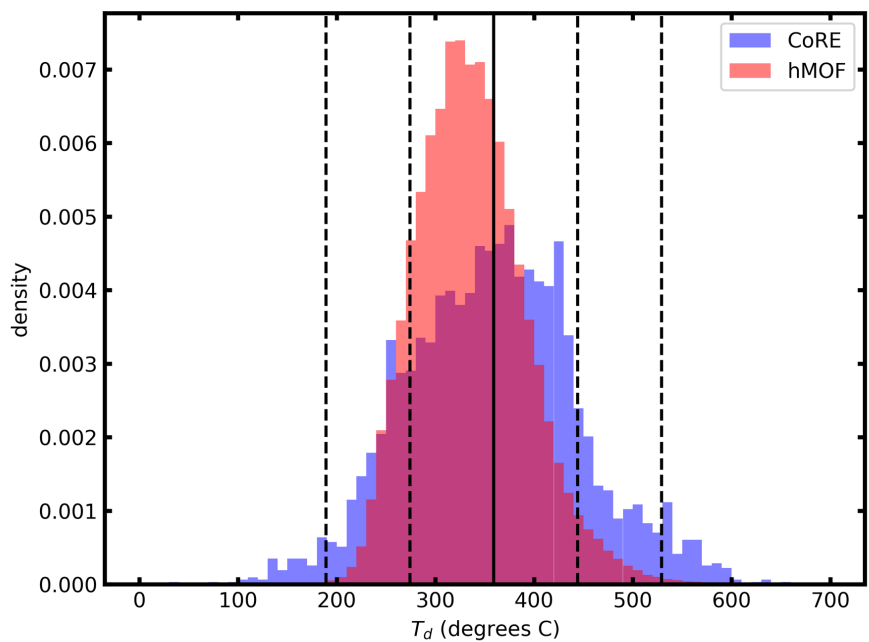

**Figure S3**. Distribution of thermal decomposition temperatures ($T_d$) as derived from CoRE MOFs (blue) and predicted on hMOFs (red). The solid line on the x-axis represents the average thermal stability of the quantified CoRE MOFs (359°C) and dashed lines represent one (87°C) and two (174°C) standard deviations above and below the average.

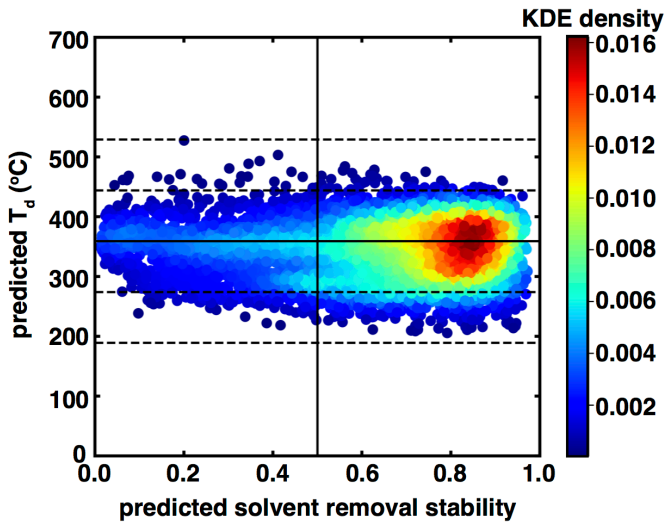

**Figure S4**. MOFs in the hMOF database that meet the uncertainty quantification (UQ) cutoffs set in prior work.[4-5] The UQ cutoffs from prior work are a latent space entropy of 0.19 for stability classification and a scaled latent space distance of 0.21 for thermal stability. The solid line on the y-axis represents the average thermal stability of the quantified CoRE MOFs (359°C) and dashed lines represent one (87°C) and two (174°C) standard deviations above and below the average. The solid line on the x-axis represents a predicted activation stability value of 0.5.

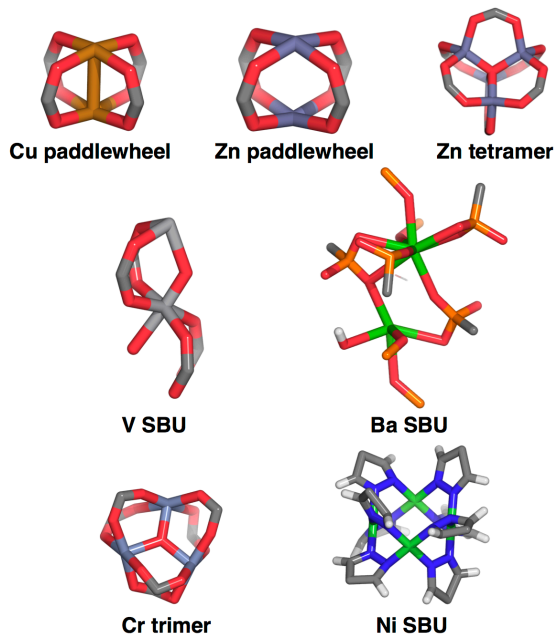

**Figure S5.** The seven inorganic nodes used for the MOFs in the BW-DB set. The atoms are colored as follows: red for O, gray for C, blue for N, orange for P, brown for Cu, dark purple for Zn, light blue for Zr, light gray for V, light purple for Cr, dark green for Ba, and light green for Ni.

**Table S2**. Counts for the structures in the BW-DB database that can be featurized and are predicted to be in the ultrastable set. We define the ultrastable set as MOFs that have a predicted activation stability value >0.5 and a predicted thermal stability that is one standard deviation (87°C) above average (359°C). Stability predictions are made on the featurizable subset of MOFs.

| step | MOFs remaining |
|---|---|
| start | 324,426 |
| featurized with RACs and geometric features | 323,718 |
| activation stability > 0.5 | 201,154 |
| thermal stability > (359+87°C) | 11,645 |
| (activation stability > 0.5) & (thermal stability > (359+87°C)) | 767 |

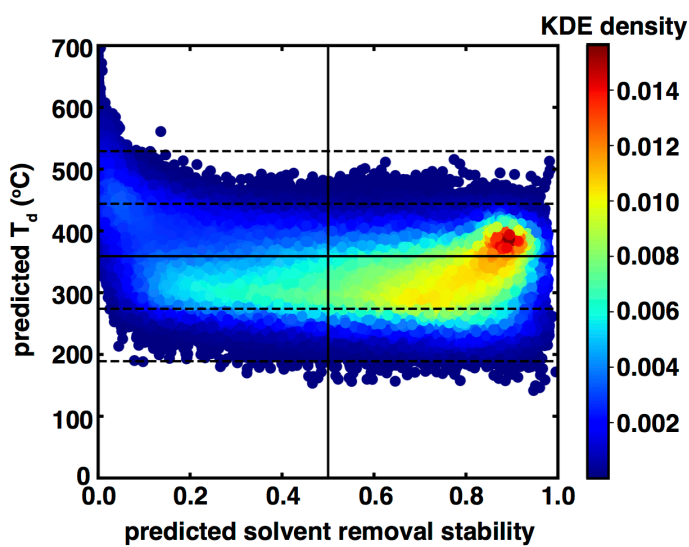

**Figure S6**. ML model predictions for MOFs in the BW-DB database.[4-5] The solid line on the y-axis represents the average thermal stability of the quantified CoRE MOFs (359°C) and dashed lines represent one (87°C) and two (174°C) standard deviations above and below average.

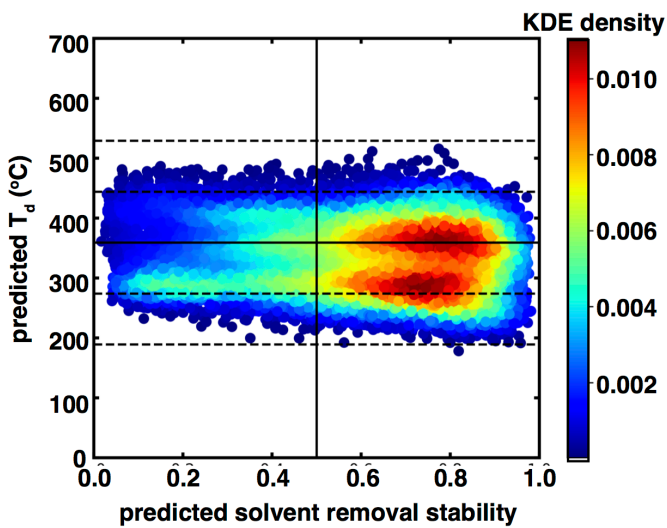

**Figure S7**. MOFs in the BW-DB database that meet the uncertainty quantification (UQ) cutoffs set in prior work.[4-5] The UQ cutoffs from prior work are a latent space entropy of 0.19 and a scaled latent space distance of 0.21. The solid line on the y-axis represents the average thermal stability of the quantified CoRE MOFs (359°C) and dashed lines represent one (87°C) and two (174°C) standard deviations above and below the average. The solid line on the x-axis represents a predicted activation stability value of 0.5. The two bands on the plot represent copper (bottom) and zinc (top) paddlewheels respectively.

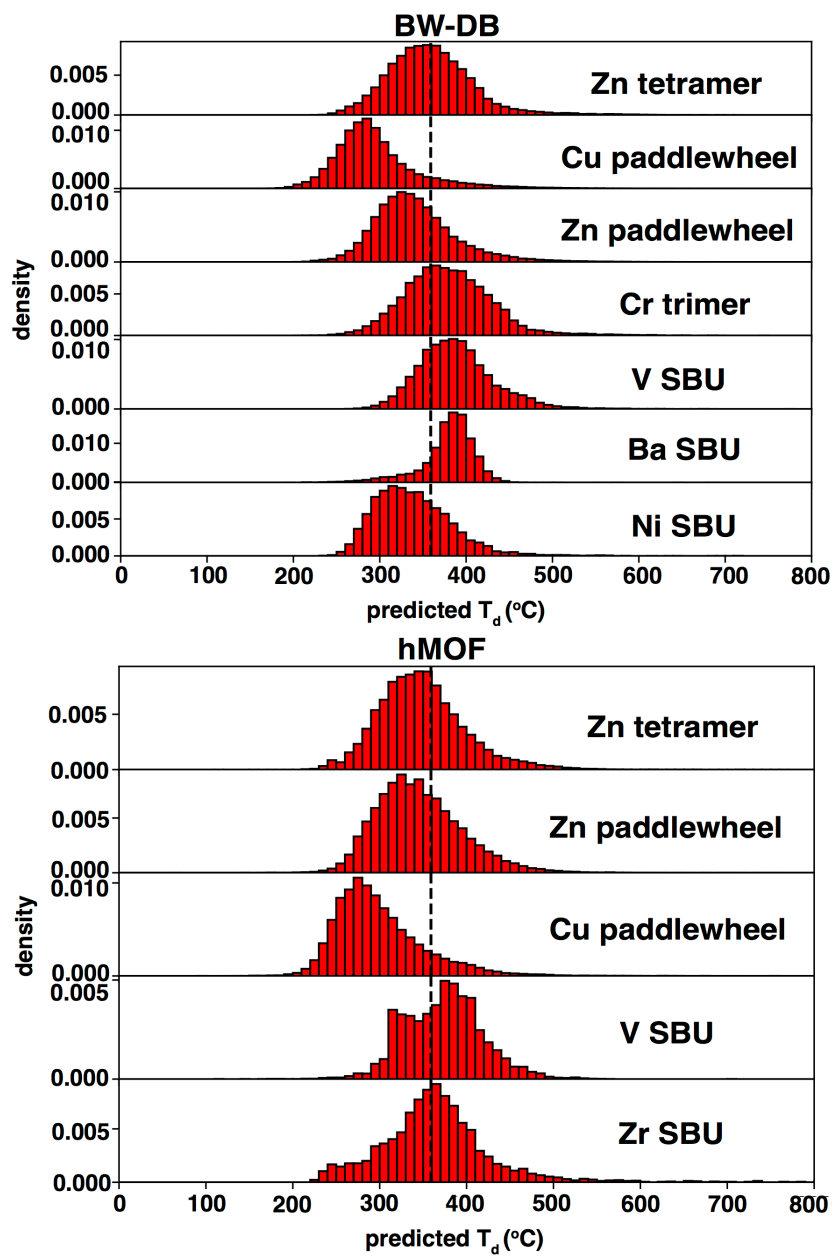

**Figure S8.** Distributions of thermal stability predictions ($T_d$) grouped by inorganic node. Predictions are made for both BW-DB (top) and hMOF (bottom) as a function of the inorganic node, as indicated inset. The average $T_d$ of CoRE MOFs is shown as a dashed line.

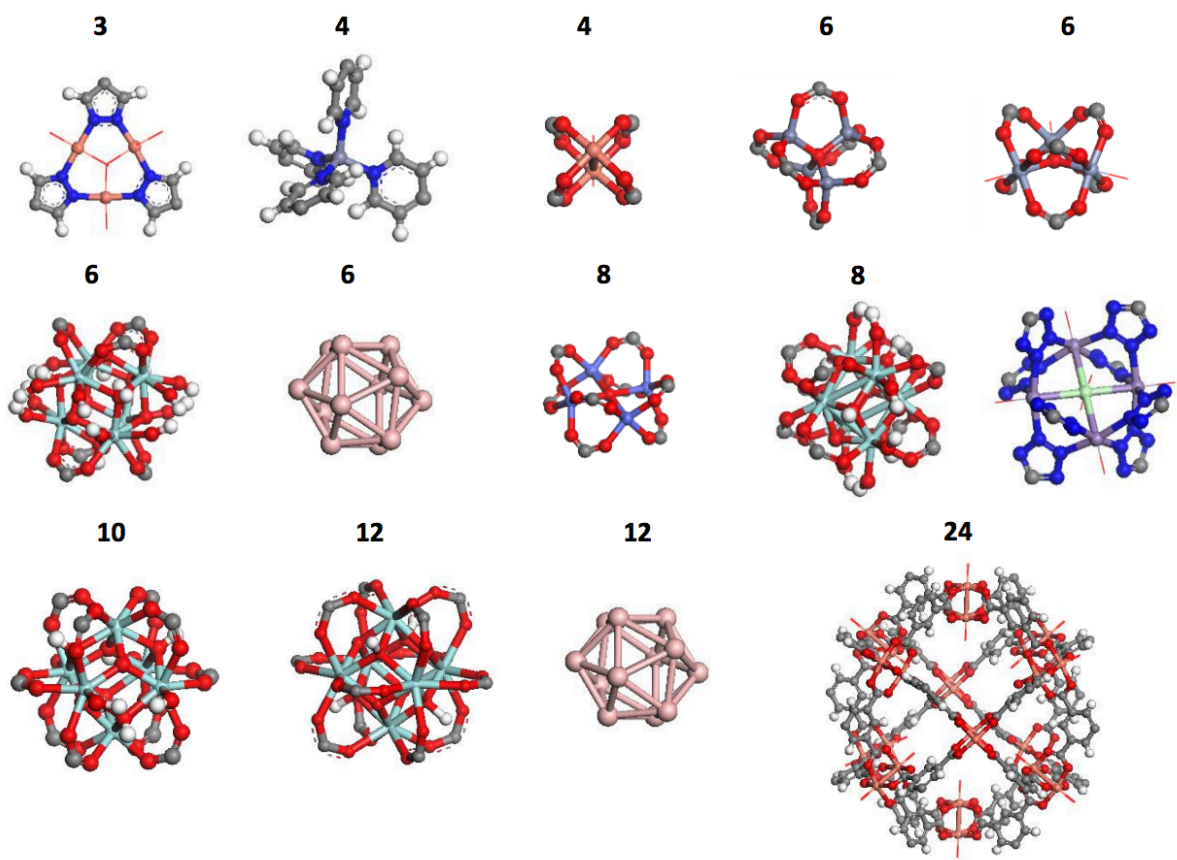

**Figure S9**. The inorganic nodes used in ToBaCCo MOFs. There are fourteen inorganic building blocks used for all MOFs. The number of coordination sites is annotated above each inorganic node. The atoms are colored as follows: red for O, gray for C, blue for N, brown for Cu, light purple for Zn, light blue for Zr, purple for Mn, light blue for Co, pink for B, light green for Cl. Figure reproduced with permission from ref.[6] Copyright 2017 American Chemical Society.

**Table S3**. Summary of the CoRE MOFs that are in the stable subset. When a ground truth is available, the ground truth (GT) is used. In other cases, a model prediction (MP) with UQ bounds from previous work (scaled LSD ≤ 0.21 for thermal stability, LSE ≤ 0.19 for activation stability) is used.[4-5] Thermal stability is assigned for MOFs with a predicted or actual $T_d$ one standard deviation (87°C) above the training data mean (359°C). Activation stability is assigned for MOFs with a predicted value above 0.5 or an actual stable assignment.

| activation stability | thermal stability | number of candidate MOFs available | number of MOFs retained after cutoffs and UQ applied |
|---|---|---|---|
| GT | GT | 1,163 | 101 |
| GT | MP | 1,016 | 13 |
| MP | GT | 1,969 | 148 |
| MP | MP | 5,420 | 122 |
| total | | | 384 |

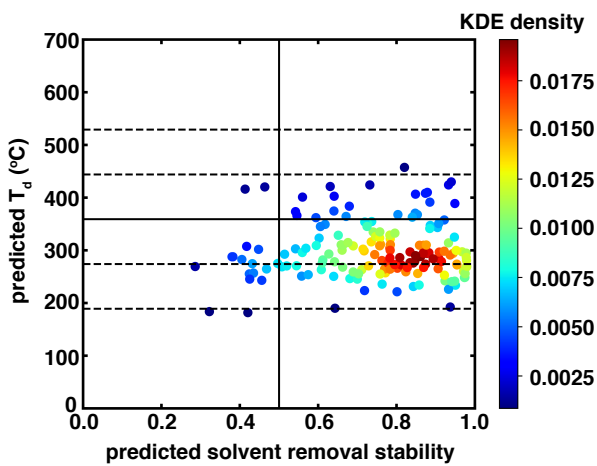

**Figure S10.** MOFs in the ToBaCCo database that meet the uncertainty quantification (UQ) cutoffs set in prior work.[4-5] The UQ cutoffs from prior work are a latent space entropy of 0.19 and a scaled latent space distance of 0.21. The solid line on the y-axis represents the average thermal stability of the quantified CoRE MOFs (359°C) and dashed lines represent one (87°C) and two (174°C) standard deviations above and below the average. The solid line on the x-axis represents a predicted activation stability value of 0.5.

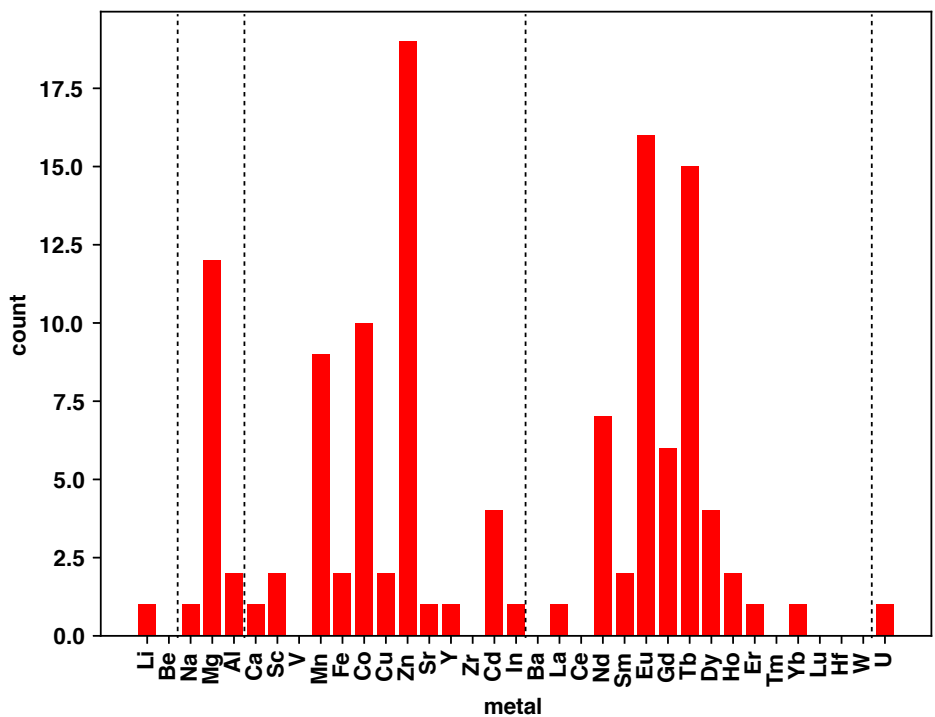

**Figure S11.** Metal counts for the SBUs in the subset of the CoRE-MOF 2019 data set from MOFs that meet the stability and uncertainty criteria (384 MOFs). Dotted black lines indicate distinct periods in the periodic table.

**Table S4**. Summary of the MOFs from the "extended" CoRE MOF internal set that are in the stable subset of MOFs. When a ground truth is available, the ground truth (GT) is used. In other cases, a model prediction (MP) with UQ bounds from previous work (scaled LSD ≤ 0.21 for thermal stability, LSE ≤ 0.19 for activation stability) is used.[4-5] Thermal stability is assigned for MOFs with a predicted or actual $T_d$ one standard deviation (87°C) above the training data mean (359°C). Activation stability is assigned for MOFs with a predicted value above 0.5 or an actual stable assignment.

| activation stability | thermal stability | number of candidate MOFs available | number of MOFs retained after cutoffs and UQ applied |
|---|---|---|---|
| GT | GT | 249 | 32 |
| GT | MP | 343 | 2 |
| MP | GT | 393 | 29 |
| MP | MP | 1,138 | 27 |
| total | | | 90 |

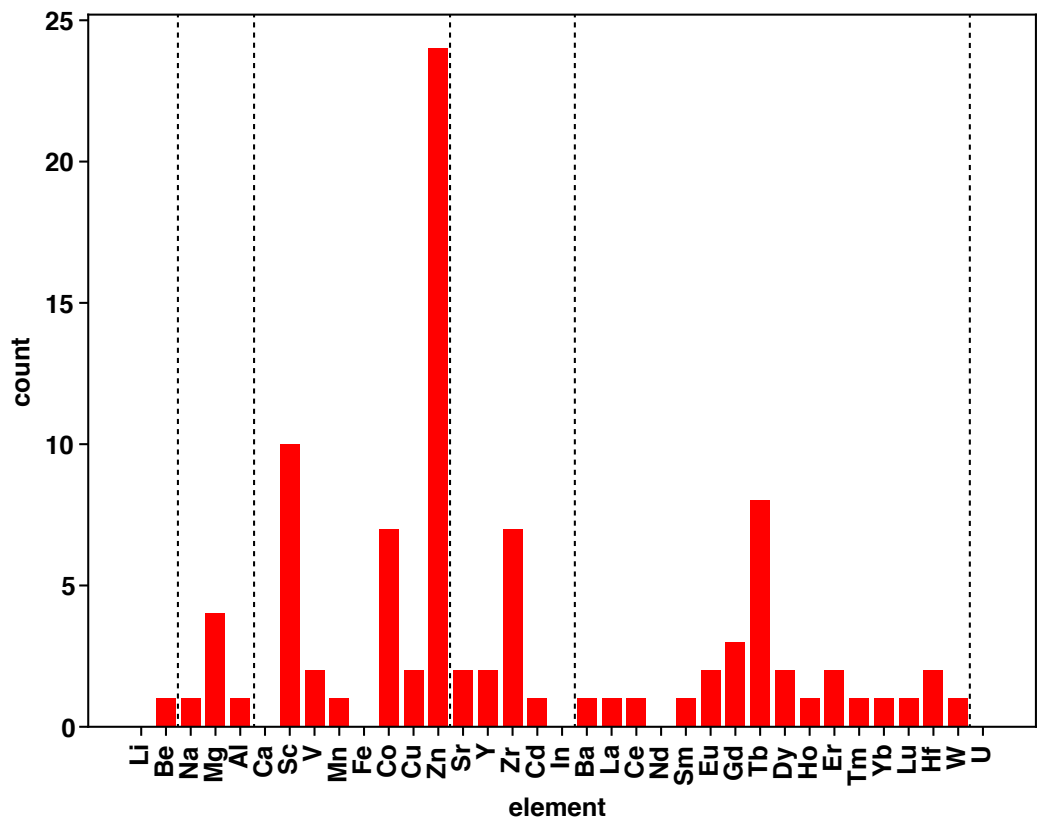

**Figure S11.** Metal counts for the SBUs in the subset of the "extended" CoRE dataset from MOFs that meet the stability and uncertainty criteria (90 MOFs). Dotted black lines indicate distinct periods in the periodic table.

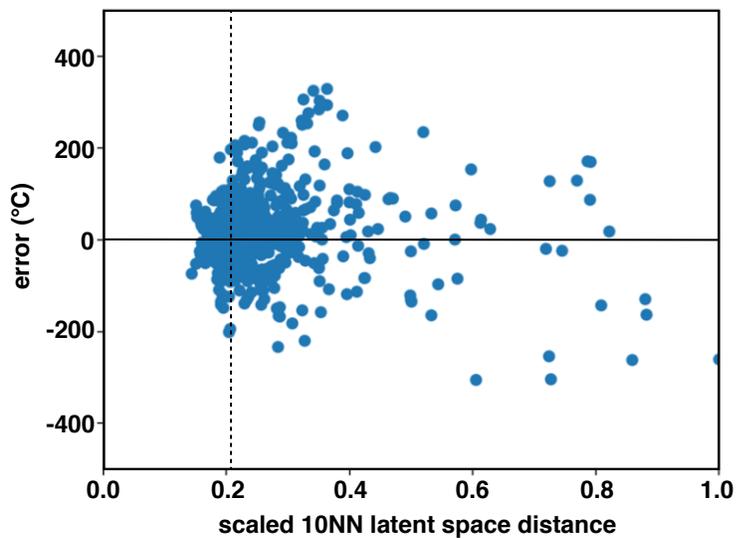

**Figure S12.** Error with respect to ground truths for thermal stability for the "extended" CoRE-MOF 2019 data set of 642 new data points vs the scaled latent space distance in the original trained model. Using a scaled latent space distance cutoff of 0.21 (shown as a vertical black dotted line) from previous work retains a mean absolute error (MAE) of 44°C, which is comparable to the model test error.[4] A solid horizontal black line corresponds to zero error.

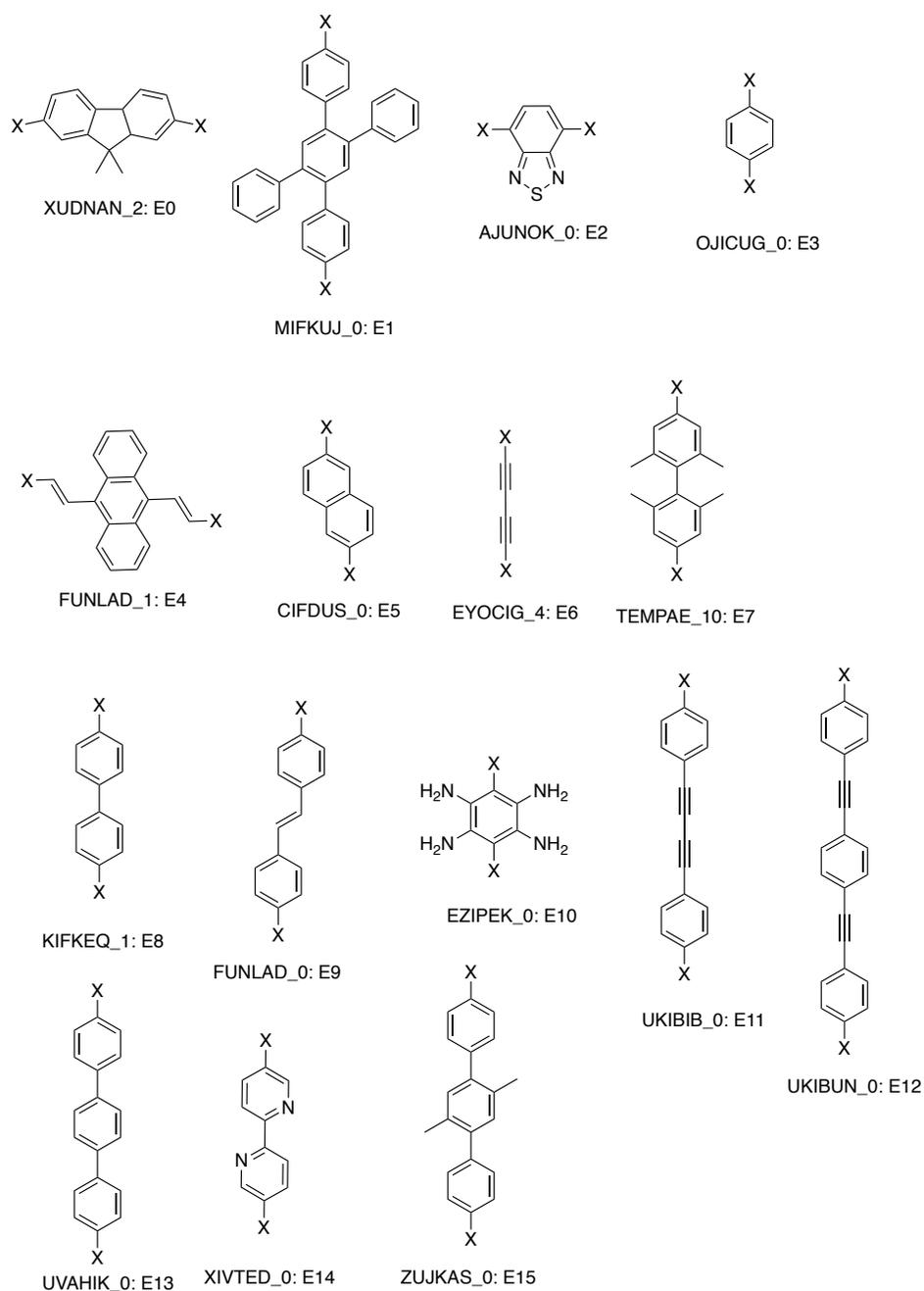

**Figure S13**. Organic edges used throughout this work. Connection points are shown as "X" atoms and the numbering for edges are given inset. For each edge, the 6-letter CSD refcode is provided along with an underscore and the number for the edge to account for multiple edges in the same MOF.

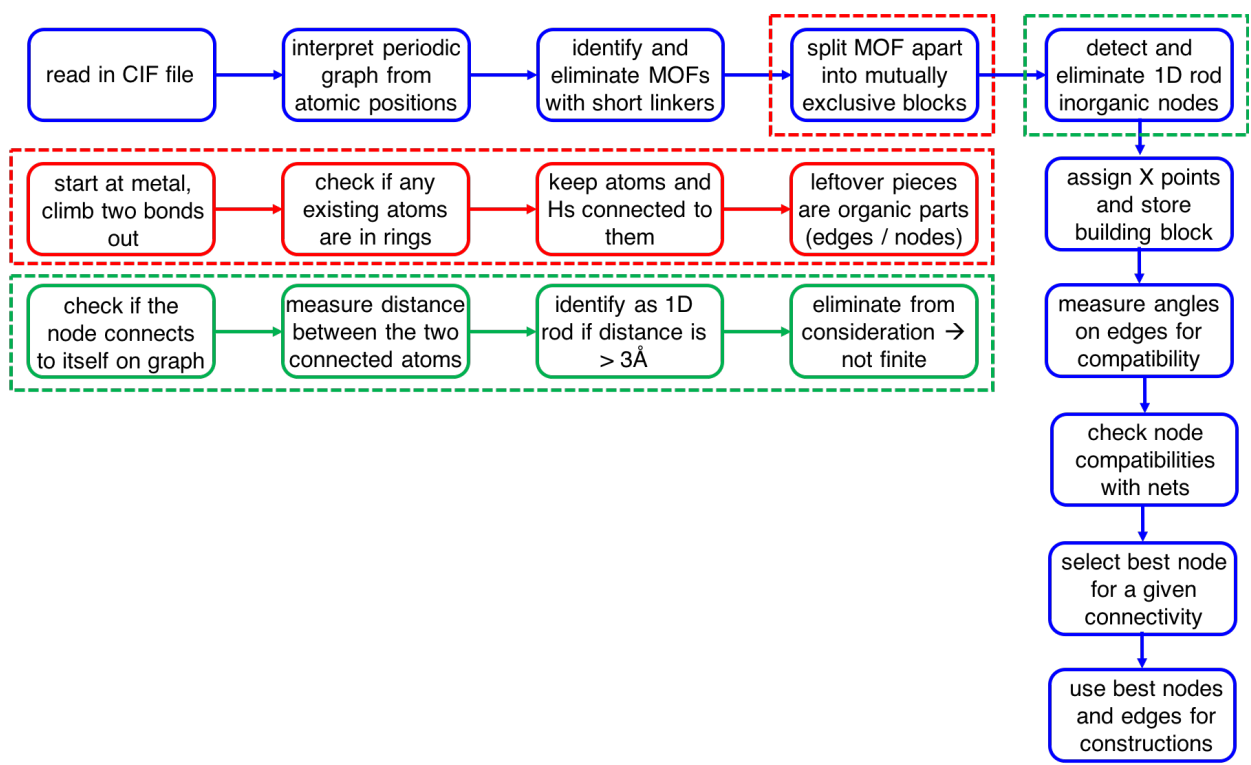

**Figure S14.** Algorithm for mutually exclusive building block extraction from existing MOFs. Every step in the process is shown in the workflow as follows: extract organic nodes, inorganic nodes, and organic edges from existing MOFs, and nodes and edges are recombined with a choice of net to construct new MOFs. All steps are shown outlined by blue boxes with arrows connecting each sequential step. For the step of splitting the MOF apart into mutually exclusive blocks, we break that step down further in the red-outlined boxes surrounded by a matching red dashed outline. For the step of detecting and eliminating 1D rod inorganic nodes, we also expand this step out into the steps in green-outlined boxes surrounded by a matching green dashed line.

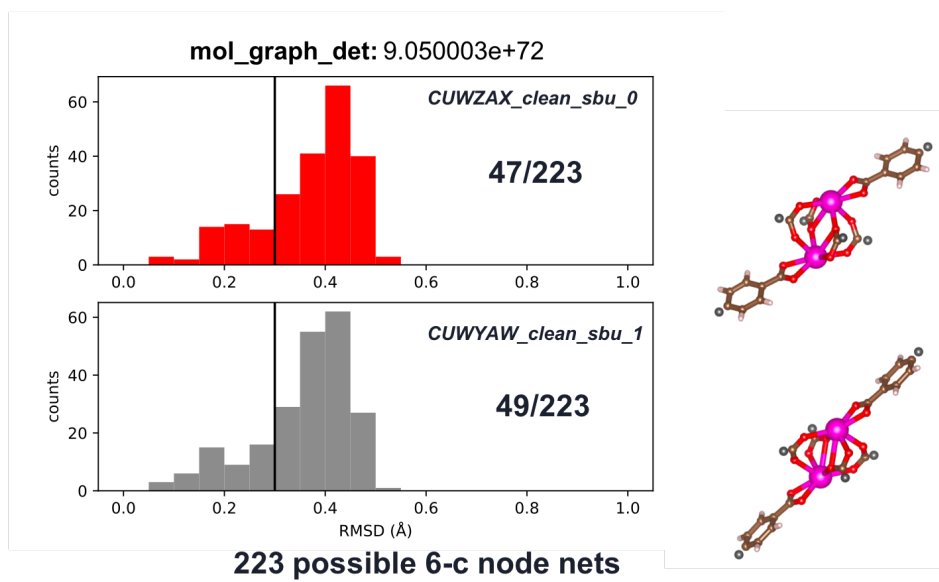

**Figure S15.** A representative example of net compatibility with a 6-connected building block. Two inorganic nodes that are the same by their molecular graph determinant are compared via their geometry to all of the 223 possible 6-connected (6-c) nets present in PORMAKE[1]. Using the 0.3-Å RMSD cutoff suggested in PORMAKE, we select the node that is compatible with more nets. The number of nets that fall within the RMSD cutoff are indicated inset as a fraction for each inorganic node.

**Table S5**. Counts for MOF construction by PORMAKE[1] and counts for successful creation of LAMMPS input files by the LAMMPS interface[7]. All MOFs were subject to a 20 minute wall time limit for optimization following a previous protocol.[3] Totals for each set are shown in the table.

| type of MOF | number constructed by PORMAKE | number prepared by LAMMPS interface | number successfully optimized | number successfully featurized |
|---|---|---|---|---|
| 1 inorganic node 1 edge | 70,444 | 53,623 | 32,491 | 32,393 |
| 1 inorganic node 1 organic node 1 edge | 27,299 | 19,399 | 15,899 | 15,896 |
| 2 inorganic nodes 1 edge | 12,252 | 9,120 | 5,854 | 5,850 |
| total | 109,995 | 82,142 | 54,244 | 54,139 |

**Table S6.** Counts for thermal and activation stability for the MOF database constructed from ultrastable parts. There are 54,139 MOFs in the database. Percentages are shown. A cutoff of 446°C for ultrastable MOFs is defined as one standard deviation (87°C) more thermally stable than average (359°C).

| condition | count (%) |
|---|---|
| thermally stable ($T_d > 359$°C) and stable upon activation | 25,536 (47%) |
| thermally unstable ($T_d \leq 359$°C) and stable upon activation | 19,342 (36%) |
| thermally stable ($T_d > 359$°C) and unstable upon activation | 4,285 (8%) |
| thermally unstable ($T_d \leq 359$°C) and unstable upon activation | 4,976 (9%) |
| ultrastable (thermally stable ($T_d > 446$°C) and stable upon activation) | 9,524 (18%) |

**Table S7**. Relative frequencies of the edges that appear in the overall data set, the ultrastable set (i.e., stable upon activation and 1 SD (87°C) more stable than average (359°C)), and the unstable cases (i.e., unstable upon activation or below average thermal stability). Absolute counts are shown at the top of each column.

| edge | overall data set $N_{MOFs} = 54,139$ | ultrastable MOFs $N_{MOFs} = 9,524$ | unstable MOFs $N_{MOFs} = 28,803$ |
|---|---|---|---|
| none | 0.079 | 0.030 | 0.110 |
| E0 | 0.074 | 0.093 | 0.058 |
| E1 | 0.069 | 0.129 | 0.041 |
| E2 | 0.085 | 0.022 | 0.106 |
| E3 | 0.084 | 0.050 | 0.086 |
| E4 | 0.016 | 0.006 | 0.022 |
| E5 | 0.080 | 0.075 | 0.070 |
| E6 | 0.015 | 0.001 | 0.025 |
| E7 | 0.073 | 0.094 | 0.058 |
| E8 | 0.085 | 0.115 | 0.071 |
| E9 | 0.017 | 0.003 | 0.025 |
| E10 | 0.072 | 0.016 | 0.107 |
| E11 | 0.001 | 0.000 | 0.001 |
| E12 | 0.001 | 0.000 | 0.001 |
| E13 | 0.084 | 0.144 | 0.070 |
| E14 | 0.086 | 0.080 | 0.088 |
| E15 | 0.080 | 0.142 | 0.062 |

**Table S8.** Simulation convergence counts for MOF mechanical properties (i.e., moduli) and deliverable capacities. Totals for each set are shown in the table.

| condition | number eliminated | number kept |
|---|---|---|
| start | - | 9,524 |
| mechanical property calculation converged | 99 | 9,425 |
| all values in stress tensor > -1 (indicates minimum) | 1,834 | 7,591 |
| moduli are positive | 261 | 7,330 |
| methane adsorption succeeds | 207 | 7,123 |

**Text S1.** Mechanical properties were computed using the formalism outlined in prior work.[8-10] We obtained the stress tensor following methods described in the main text (see Methods). Upon calculation of the stress tensor, we obtained both the bulk elastic and shear moduli using the Voight (V) and Reuss (R) methods to get upper and lower bounds. We then obtained the Hill (H) modulus by averaging the Voight and Reuss moduli. After obtaining the stress tensor (C), we computed the Voight and Reuss elastic (K) and shear (G) moduli as follows. We compute the inverse of the stress tensor as the stiffness tensor S:

$$S = C^{-1}$$

$$K_R = \frac{1}{S_{11} + S_{22} + S_{33} + 2*(S_{12} + S_{13} + S_{23})}$$

$$K_V = \frac{C_{11} + C_{22} + C_{33} + 2*(C_{12} + C_{13} + C_{23})}{9.0}$$

$$K_{VRH} = \frac{(K_V + K_R)}{2}$$

$$G_R = \frac{15}{(4*(S_{11} + S_{22} + S_{33}) - 4*(S_{12} + S_{13} + S_{23}) + 3*(S_{44} + S_{55} + S_{66}))}$$

$$G_V = \frac{(C_{11} + C_{22} + C_{33}) - (C_{12} + C_{13} + C_{23}) + 3*(C_{44} + C_{55} + C_{66})}{15}$$

$$G_{VRH} = \frac{(G_V + G_R)}{2}$$

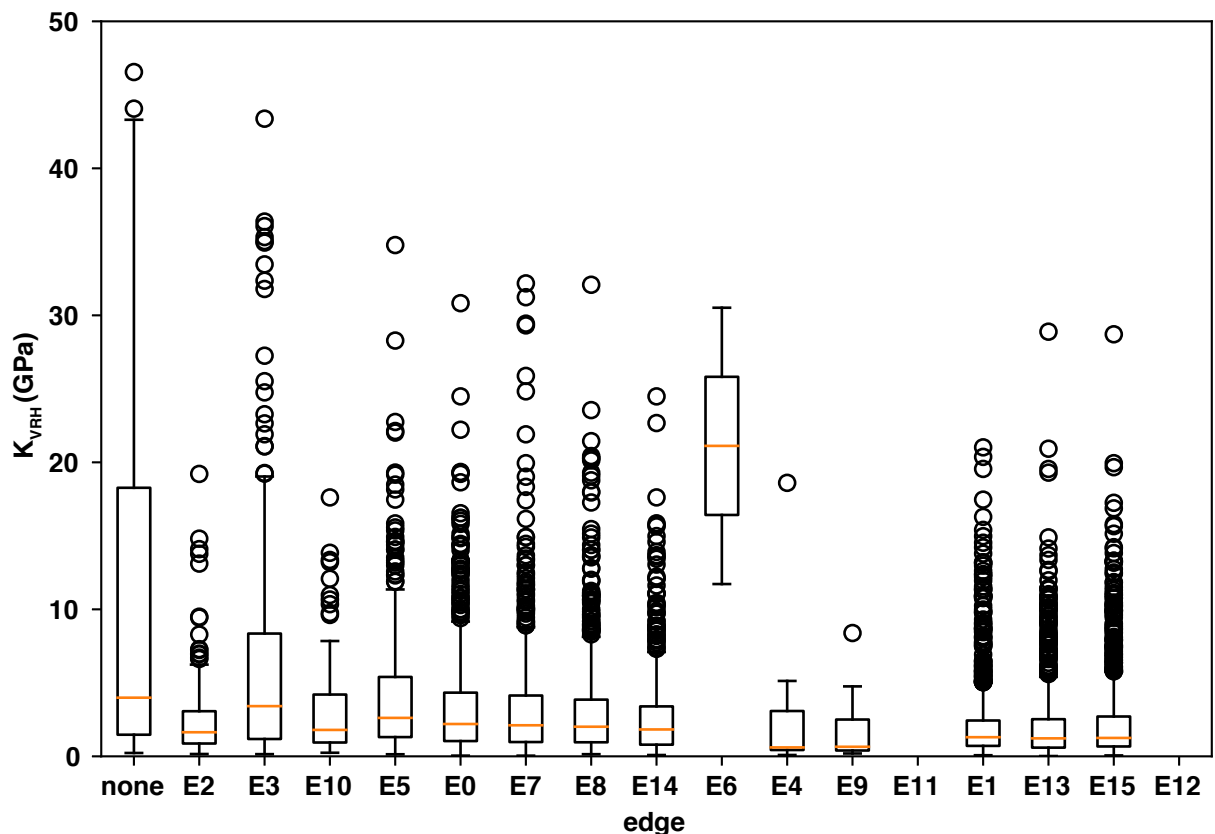

**Figure S16.** Box and whisker plots of the edge dependence of the bulk elastic modulus. We order the edges by the distance between X atoms, which indicates edge length. E11 and E12 do not have any converged results, and E6 (marked with an asterisk) only contains two converged results. The box represents quartile 1 (lower) and quartile 3 (upper) with the median marked in orange. The whiskers extend to 1.5 times the interquartile range above quartile 3 and below quartile 1. Dots represent all points outside of the whiskers. The y-axis is truncated at 50 GPa, eliminating five outlier points for the "none" category that are not shown.

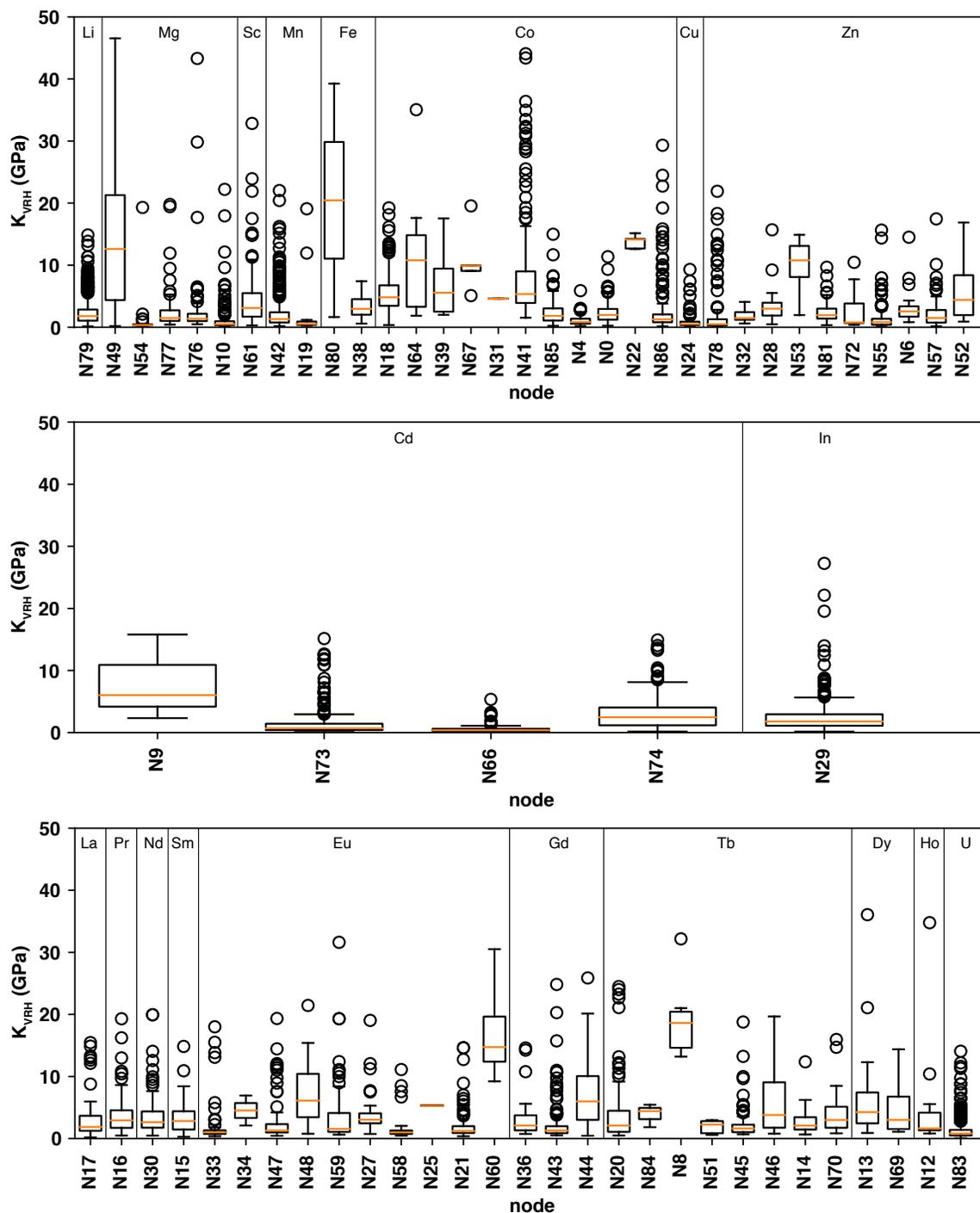

**Figure S17.** Box and whisker plots of the node dependence of the bulk elastic modulus. We group the nodes by metal identity in the periodic table as indicated in inset labels between each set of vertical lines. The box represents quartile 1 (lower) and quartile 3 (upper) with the median marked in orange. The whiskers extend to 1.5 times the interquartile range above quartile 3 and below quartile 1. Dots represent all points outside of the whiskers. The y-axis is truncated at 50 GPa, which truncates the plot and leads to one outlier point omitted for each of the following nodes: N49, N41, N47, N48, and N45.

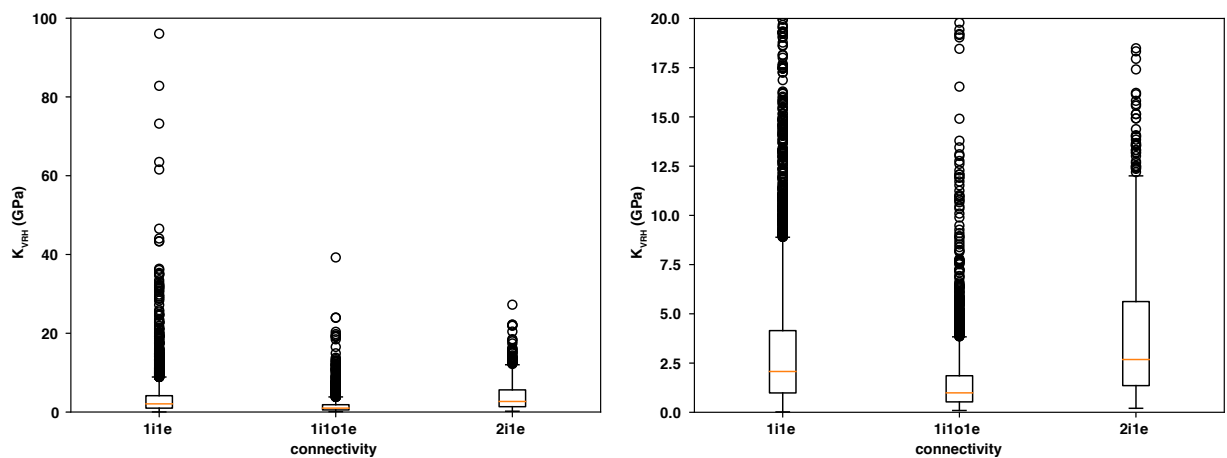

**Figure S18.** Box and whisker plots of the elastic modulus for different MOF connectivities. We label MOFs constructed from 1 inorganic node and 1 edge as 1i1e; 1 inorganic node, 1 organic node, and 1 edge as 1i1o1e; and 2 inorganic nodes and 1 edge as 2i1e. The same data with a truncated y-axis range is shown on the right. The box represents quartile 1 (lower) and quartile 3 (upper) with the median marked in orange. The whiskers extend to 1.5 times the interquartile range above quartile 3 and below quartile 1. Dots represent all points outside of the whiskers.

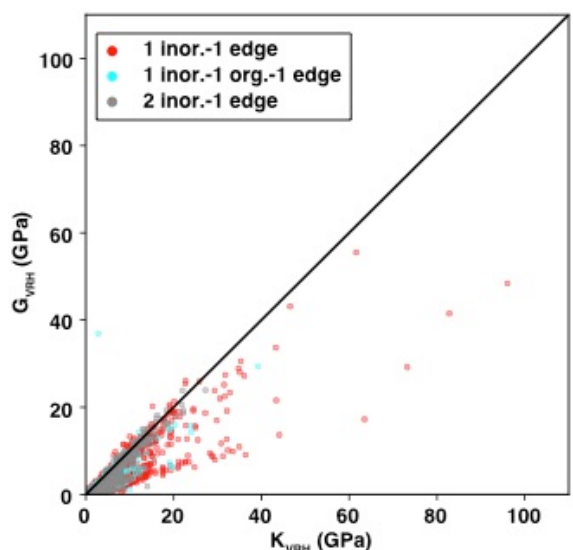

**Figure S19.** Comparison of bulk elastic and shear moduli by MOF connectivity type. Both elastic ($K_{VRH}$) and shear ($G_{VRH}$) moduli are shown as the Hill averages. Pearson correlations between elastic and shear moduli are 0.85 for one inorganic node and one edge; 0.89 for one inorganic node, one organic node, and one edge; and 0.92 for two inorganic nodes and one edge. The overall Pearson correlation for all data is 0.88. Data points are represented as translucent circles to depict data density.

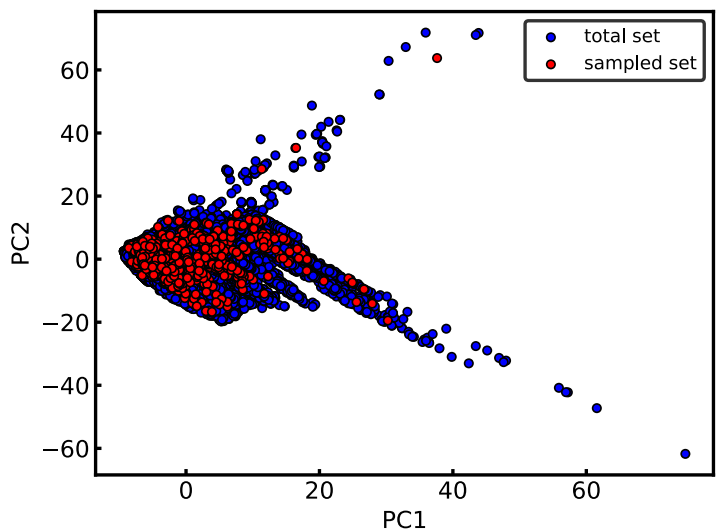

**Figure S20.** Principal component analysis (PCA) plot of total unstable set (blue) and sampled subset of unstable MOFs (red). 1,000 MOFs are sampled from the set of 44,625 unstable MOFs using k-medoids. 1,000 k-medoids clusters were created on revised autocorrelation (RAC) features and geometric features. The calculation used the Euclidean distance metric and alternate algorithm. The random seed was 2022.

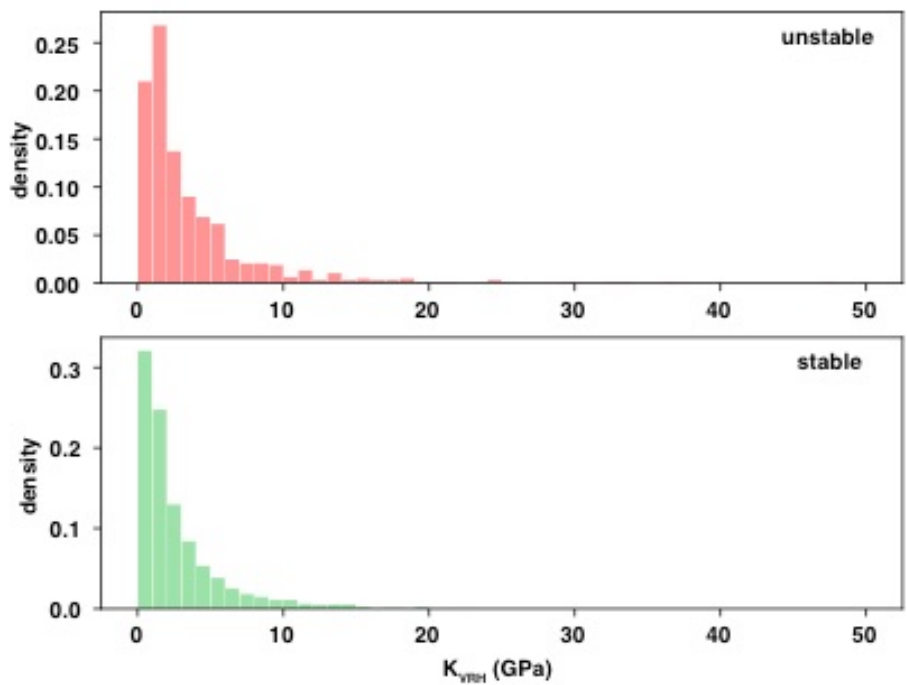

**Figure S21** Histograms of elastic moduli for unstable (top) and stable (bottom) MOF subsets. All elastic moduli are shown as Voight–Reuss–Hill averages.

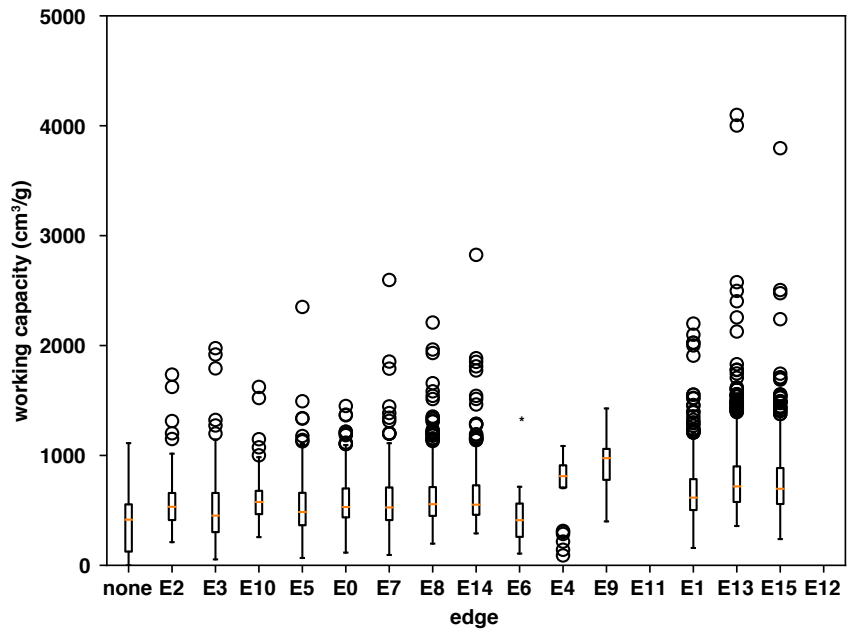

**Figure S22.** Box and whisker plots showing the edge dependence of CH$_4$ deliverable capacity. We order the edges by distance between X atoms, which indicates edge length. E11 and E12 do not have any converged results, and E6 (marked with an asterisk) only contains two converged results. The box represents quartile 1 (lower) and quartile 3 (upper) with the median marked in orange. The whiskers extend to 1.5 times the interquartile range above quartile 3 and below quartile 1. Dots represent all points outside of the whiskers.

**Table S9.** Deliverable capacities and mechanical properties for Pareto-optimal MOFs from the stable subset for which all properties were computed. The type of MOF (1 inorganic node-1 edge: 1i1e, 1 inorganic node-1 organic node-1 edge: 1i1o1e, 2 inorganic nodes-1 edge: 2i1e), the net, the inorganic node, the organic node, the edge, and the corresponding properties are also listed in the table.

| type / net | inorganic node, metal | organic node | edge | CH$_4$ deliverable capacity, cm$^3$/g | elastic modulus (K$_{VRH}$), GPa |
|---|---|---|---|---|---|
| 1i1e / qtz-e | N45, Tb | - | none | 0 | 96.1 |
| 1i1e / qtz-e | N47, Eu | - | none | 0.002 | 82.8 |
| 1i1e / uon | N41, Co | - | none | 124.0 | 63.5 |
| 1i1e / ulj | N41, Co | - | none | 283.8 | 44.1 |
| 1i1e / ukj | N41, Co | - | E3 | 443.8 | 36.4 |
| 1i1e / vba | N41, Co | - | E3 | 446.6 | 35.0 |
| 1i1e / vbb | N41, Co | - | E3 | 515.7 | 32.4 |
| 1i1e / ukj | N41, Co | - | E8 | 597.4 | 32.1 |
| 1i1e / vbb | N41, Co | - | E7 | 673.5 | 29.4 |
| 1i1e / ukj | N41, Co | - | E13 | 767.7 | 28.9 |
| 1i1e / ukk | N41, Co | - | E13 | 782.8 | 20.9 |
| 1i1e / uml | N54, Mg | - | E13 | 1141.4 | 19.3 |
| 1i1o1e / ptr | N18, Co | orgN19 | E1 | 1231.7 | 5.9 |
| 1i1o1e / ssa | N41, Co | orgN18 | E13 | 1235.8 | 2.8 |
| 1i1o1e / ctn | N10, Mg | orgN14 | E15 | 1366.6 | 2.6 |
| 1i1o1e / bpr | N83, U | orgN23 | E13 | 1376.4 | 1.9 |
| 1i1o1e / ssa | N41, Co | orgN19 | E13 | 1394.3 | 1,9 |
| 1i1o1e / ptr | N79, Li | orgN21 | E13 | 1470.4 | 1.8 |
| 1i1o1e / ptr | N74, Cd | orgN19 | E13 | 1487.4 | 1.5 |
| 1i1e / sod-g | N83, U | - | E3 | 1792.1 | 1.5 |
| 1i1e / etb | N83, U | - | E1 | 2098.4 | 0,9 |
| 1i1e / srs-a | N83, U | - | E5 | 2351.0 | 0.7 |
| 1i1e / srs-a | N83, U | - | E7 | 2596.7 | 0.7 |
| 1i1e / sod-h | N83, U | - | E13 | 4002.0 | 0.1 |
| 1i1e / 1cv-f | N66, Cd | - | E13 | 4098.6 | 0.1 |

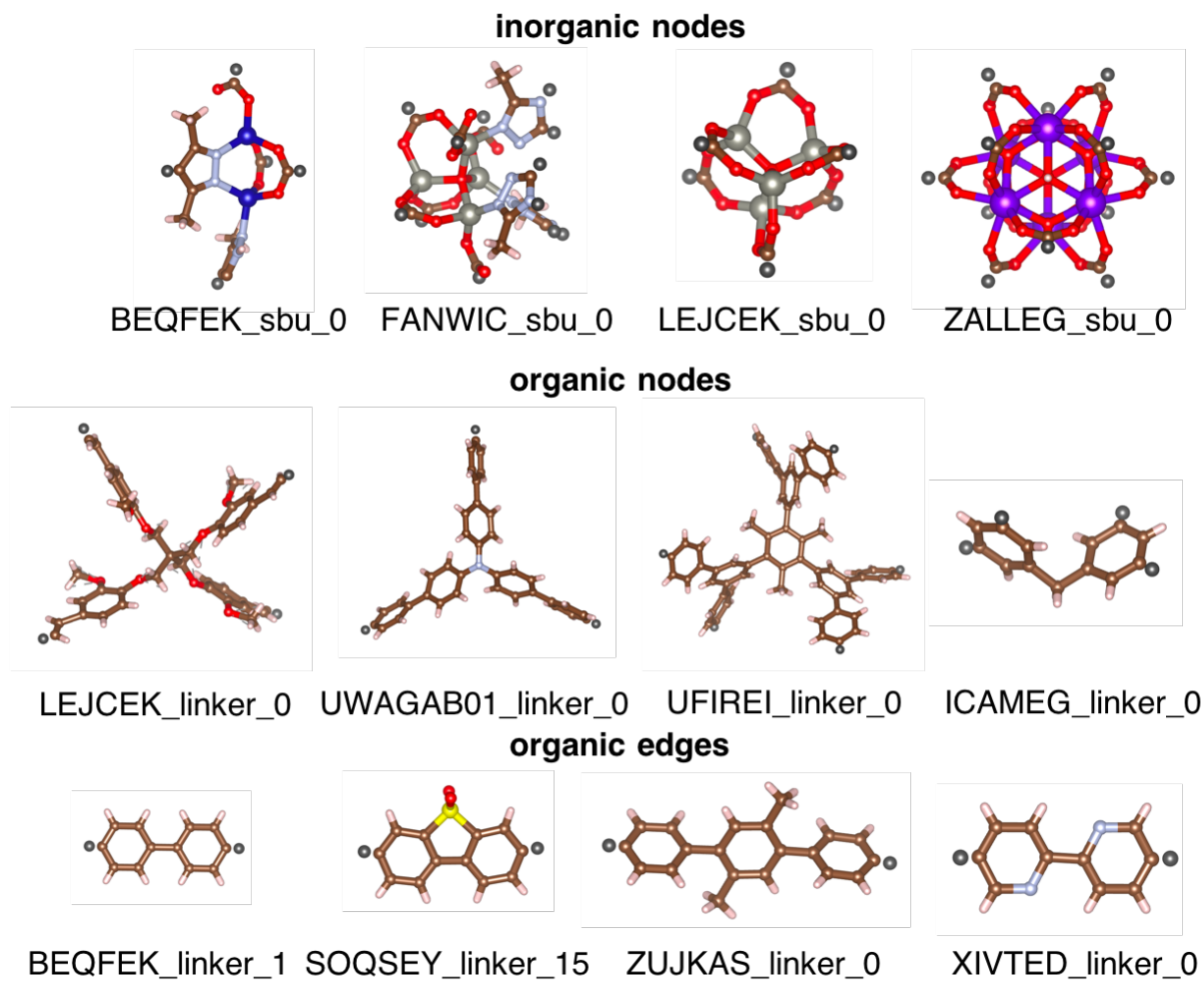

**inorganic nodes**

BEQFEK_sbu_0    FANWIC_sbu_0    LEJCEK_sbu_0    ZALLEG_sbu_0

**organic nodes**

LEJCEK_linker_0    UWAGAB01_linker_0    UFIREI_linker_0    ICAMEG_linker_0

**organic edges**

BEQFEK_linker_1    SOQSEY_linker_15    ZUJKAS_linker_0    XIVTED_linker_0

**Figure S23.** Representative building blocks extracted from existing experimental MOFs labeled by their refcode in the CSD and whether they are an SBU or linker. One MOF may have multiple building blocks extracted. The final number in the label identifies which building block number (zero indexed) was extracted from the MOF. Inorganic and organic nodes refer to building blocks with > 2 connection points, where the former contains metals while the latter contains only main group elements. Edges are exclusively organic and contain only two connection points. All connection points are shown as black "X" atoms. Atoms are colored as follows: blue for Co, gray for Zn, purple for Tb, brown for C, red for O, light blue for N, yellow for S, white for H, and black for X. All extracted and utilized building blocks are reported in the Supporting Information.

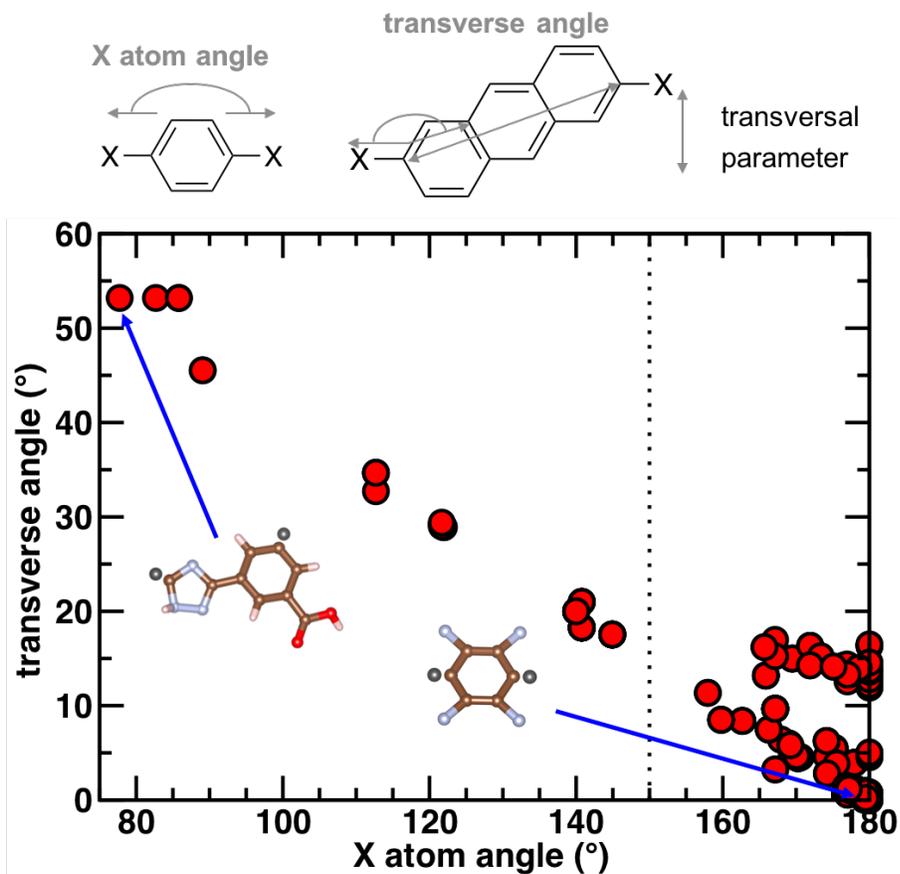

**Figure S24.** The X atom angle (in degrees), which is the angle between the connection angle vectors (e.g. vectors of the C–X bonds), versus the transverse angle (in degrees), which represents the angle between a vector along the C–X bond of one X-connection with the angle between the C of the neighboring X atom, as shown at top. A cutoff of 150° for the X-atom angles selected by trial and error is shown as a dotted black vertical line. This also in practice bounds the transverse angle. Example edges are shown inset, with C in brown, N in light blue, O in red, H in white, and X in black.

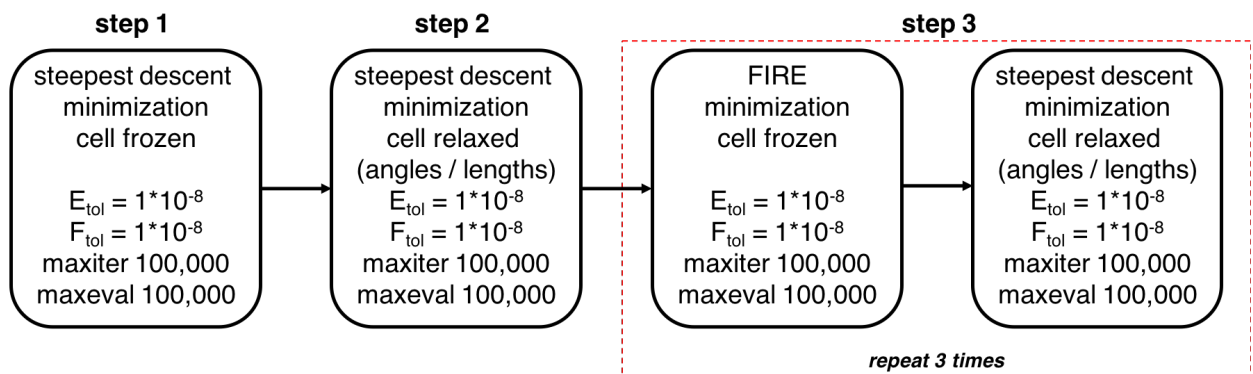

**Figure S25.** Step-by-step workflow for optimizing MOFs after constructing them from PORMAKE.[1] After construction, MOFs are iteratively optimized using the steepest descent and FIRE algorithms as implemented in LAMMPS.[2] All convergence criteria are shown for each step. This workflow follows a previously proposed workflow to optimize covalent organic framework (COF) structures.[3] "Cell relaxed" corresponds to steps in which both cell angles and cell lengths can change. Atomic coordinates are permitted to relax at all steps in the process.

**Table S10.** Description of revised autocorrelation (RAC) features used in this work with the start / scope, operation performed and maximum depth, atomwise properties autocorrelated, features removed, and total feature count. We use five heuristic atomwise properties for all operations: nuclear charge (Z), topology (T), identity (I), electronegativity ($\chi$), and covalent radius (S). RAC features use the notation <start>-<atomic property>-<depth>-<scope>. Difference RACs contain a "D" prefix with a subscripted start.

| start / scope | operation, maximum depth | atomwise quantities | features removed | feature count |
|---|---|---|---|---|
| metal centered / all | product, 3 | Z, T, I, $\chi$, S | 1 (mc-I-0-all) | 19 |
| metal centered / all | difference, 3 | Z, T, I, $\chi$, S | 8 ($D_{mc}$-I-0-all, $D_{mc}$-I-1-all, $D_{mc}$-I-2-all, $D_{mc}$-I-3-all, $D_{mc}$-S-0-all, $D_{mc}$-T-0-all, $D_{mc}$-Z-0-all, $D_{mc}$-$\chi$-0-all) | 12 |
| linker connecting / linker | product, 3 | Z, T, I, $\chi$, S | 1 (lc-I-0-linker) | 19 |
| linker connecting / linker | difference, 3 | Z, T, I, $\chi$, S | 8 ($D_{lc}$-I-0-linker, $D_{lc}$-I-1-linker, $D_{lc}$-I-2-linker, $D_{lc}$-I-3-linker, $D_{lc}$-S-0-linker, $D_{lc}$-T-0-linker, $D_{lc}$-Z-0-linker, $D_{lc}$-$\chi$-0-linker) | 12 |
| functional group / linker | product, 3 | Z, T, I, $\chi$, S |  | 20 |
| functional group / linker | difference, 3 | Z, T, I, $\chi$, S | 8 ($D_{func}$-I-0-linker, $D_{func}$-I-1-linker, $D_{func}$-I-2-linker, $D_{func}$-I-3-linker, $D_{func}$-S-0-linker, $D_{func}$-T-0-linker, $D_{func}$-Z-0-linker, $D_{func}$-$\chi$-0-linker) | 12 |
| full unit cell / all | product, 3 | Z, T, I, $\chi$, S | 0 | 20 |
| full linker / linker | product, 3 | Z, T, I, $\chi$, S | 0 | 20 |
|  |  |  | 26 | 134 |

**Table S11.** Description of geometric features generated by Zeo++ with definitions and units. We used a nitrogen probe molecule with a radius of 1.86 Å to generate all geometric features. There are 14 geometric features in total.

| feature | meaning | units |
|---------|---------|-------|
| $D_f$ | maximum free sphere | Å |
| $D_i$ | maximum included sphere | Å |
| $D_{if}$ | maximum included sphere in free sphere path | Å |
| GPOAV | gravimetric pore accessible volume | cm³/g |
| GPONAV | gravimetric pore non- accessible volume | cm³/g |
| GPOV | gravimetric pore volume | cm³/g |
| GSA | gravimetric surface area | m²/g |
| POAV | pore accessible volume | Å³ |
| PONAV | pore non-accessible volume | Å³ |
| POAVF | pore accessible volume fraction | unitless |
| PONAVF | pore non-accessible volume fraction | unitless |
| VPOV | volumetric pore volume | cm³/cm³ |
| VSA | volumetric surface area | m²/cm³ |
| $\rho$ | crystal density | g/cm³ |